\documentclass[12pt]{article}
\usepackage{epsfig}
\usepackage{xspace}
\usepackage{amssymb,amsmath,amsbsy}
\usepackage{mathrsfs}
\usepackage{dsfont}
\usepackage{geometry}  
\usepackage{graphicx}
\usepackage{booktabs}
\usepackage{cite}

\def\theequation{\arabic{section}.\arabic{equation}}

     {\arraycolsep 0.24em\begin{eqnarray}}{\end{eqnarray}}
\def\beq{\begin{equation}}
\def\eeq{\end{equation}}
\def\slashchar#1{\setbox0=\hbox{$#1$}           
   \dimen0=\wd0                                 
   \setbox1=\hbox{/} \dimen1=\wd1               
   \ifdim\dimen0>\dimen1                        
      \rlap{\hbox to \dimen0{\hfil/\hfil}}      
      #1                                        
   \else                                        
      \rlap{\hbox to \dimen1{\hfil$#1$\hfil}}   
      /                                         
   \fi}                                        %

\def\eq#1{eq.~(\ref{#1})}
\def\Eq#1{Eq.~(\ref{#1})}

\def\eqs#1#2{eqs.~(\ref{#1}) and (\ref{#2})}

\def\Ref#1{ref.~\cite{#1}}

\def\Refs#1{refs.~\cite{#1}}

\newcommand{\ETmiss}{\mbox{\slash \hspace*{-0.25cm}$E_{T}$}\xspace}
\newcommand{\Emiss}{\mbox{\slash \hspace*{-0.25cm}$E$}\xspace}
\newcommand{\pTmiss}{\mbox{\slash \hspace*{-0.2cm}$p_{T}$}\xspace}
\newcommand{\Sherpa}{S\scalebox{0.8}{HERPA}\xspace} 
\newcommand{\Pythia}{P\scalebox{0.8}{YTHIA}\xspace} 
\newcommand{\htohv}{h\scalebox{0.8}{2hv}\xspace} 

\begin{document}
\tolerance=100000

\setcounter{page}{0}
\thispagestyle{empty}

\begin{flushright}
IPPP-08-56\\[-1mm]
CPT-08-112\\[-1mm]
\end{flushright}

\bigskip

\begin{center}
{\Large \bf Searching for Nambu - Goldstone Bosons at the LHC }\\[1.7cm]


{{\large Athanasios Dedes}$^{\, a}$, {\large Terrance Figy}$^{\, b}$,
{\large Stefan H{\"o}che}$^{\, b}$, \\[4mm]{\large Frank Krauss}$^{\, b}$, and 
{\large Thomas ~E.~J. ~Underwood}$^{\,c}$}\\[0.5cm]
{\it $^a$Division of Theoretical Physics, University of Ioannina, Ioannina,
GR 45110,  Greece}\\[3mm]
{\it $^b$Institute for Particle Physics Phenomenology, University of
Durham, DH1 3LE, UK}\\[3mm]
{\it $^c$Max-Planck-Institut f\"ur Kernphysik, Saupfercheckweg 1, 69117
Heidelberg, Germany}\\[3mm]
\end{center}

\date{\today}

\vspace*{0.8cm}\centerline{\bf ABSTRACT}
\vspace{0.5cm}
\noindent{\small

Phenomenological implications of a minimal extension to the Standard Model are considered, 
in which a Nambu-Goldstone boson emerges from the spontaneous breaking of a global $U(1)$ 
symmetry.  This is felt only by a scalar field which is a singlet under all Standard Model 
symmetries, and possibly by neutrinos.  Mixing between the Standard Model Higgs boson 
field and the new singlet field may lead to predominantly invisible Higgs boson decays.  
The ``natural'' region in the Higgs boson mass spectrum is determined, where this minimally 
extended Standard Model is a valid theory up to a high scale related with the smallness of 
neutrino masses.  Surprisingly, this region may coincide with low visibility of {\em all} 
Higgs bosons at the LHC.  Monte-Carlo simulation studies of this ``nightmare'' situation 
are performed and strategies to search for such Higgs boson to invisible (Nambu-Goldstone 
boson) decays are discussed.  It is possible to improve the signal-to-background ratio 
by looking at the distribution of either the total transverse momentum of the leptons and 
the \pTmiss, or by looking at the distribution of the azimuthal angle between the \pTmiss
and the momentum of the lepton pair for the $Z$- and Higgs-boson associated production.  
We also study variations of the model with non-Abelian symmetries and present approximate 
formulae for Higgs boson decay rates.  Searching for Higgs bosons in such a scenario at the 
LHC would most likely be solely based on Higgs to ``invisible'' decays.
}

\vspace*{\fill}
\newpage

\setcounter{equation}{0}
\section{Introduction}

In a seminal paper published in 1962, Goldstone, Salam and Weinberg~\cite{GSW}
proved that the physical particle spectrum of a theory in which a continuous,
global symmetry is spontaneously broken must contain one massless, spin-zero
particle for each broken symmetry. Massless particles of this type, today
called Nambu-Goldstone bosons (NGB), were first theoretically discovered in
particular models by Goldstone~\cite{Goldstone} and Nambu~\cite{Nambu}.  In
the following, they will collectively be denoted by the symbol $\mathcal{J}$.
NGBs have the peculiar property that they couple to the divergence of the
current $j^{\mu}(x)$ associated with the symmetry that is broken.  This coupling 
has a strength which is inversely proportional to the scale of symmetry 
breaking $F$,
\begin{eqnarray}
\mathcal{L}_{\rm int} \ = \ 
\frac{1}{2\: F} \: \mathcal{J}(x) \cdot  \partial_{\mu} j^{\mu}(x) \;.
\label{eq1}
\end{eqnarray}
This form of interaction is invariant under the shift transformation,
$\mathcal{J} \rightarrow \mathcal{J} + \omega$, where $\omega$ is an angle that 
parameterizes different vacuum field configurations.  Since NGBs typically are 
amongst the lightest particles in a theory a large fraction of the other particles 
can decay into them through \eq{eq1}.  For this decay to occur, these other particles,
possibly scalars and/or quarks and leptons, must be charged under the same
spontaneously broken global symmetry.  It was first proposed by Suzuki and
Schrock~\cite{SS} that if the Standard Model (SM) Higgs boson mixes with such
a new scalar particle then it must have a decay channel into a pair of NGBs
($\mathcal{J}\mathcal{J}$) provided that the scale $F$ is of the order of the
electroweak gauge boson masses, $F \approx 100~\mathrm{GeV}$.  If such a Higgs
boson decay exists then it should be searched for at colliders.

The basic idea underlying this article is the existence of an additional
global ``phantom'' symmetry, $G_{\rm P}=U(1)_{\rm P}$ (P stands for
``Phantom''), that is spontaneously broken at some scale $F$.  Then following
\eq{eq1}, $\mathcal{J}$ will couple to all fermions ($f$) that are charged
under $G_{\rm P}$ since $\partial_{\mu} j^{\mu} = m_{f} \bar{f}\gamma_{5} f$.
This coupling will be proportional to $m_{f}/F$.  In the literature, there are
three famous types of Nambu-Goldstone bosons: axions~\cite{Axion},
familons~\cite{Wilczek} and majorons~\cite{Moh} and their associated broken
symmetries are the Peccei-Quinn symmetry~\cite{PQ}, and the family and lepton
number symmetry, respectively.  In the former two cases the global symmetry is
carried by both quarks and leptons and in the latter case by leptons only.
However, considerations of energy loss in stars, supernovae and/or in
terrestrial collider experiments~\cite{PDG} conclude that $F \gtrsim
10^{9-10}$ GeV in these popular cases.  This bound constrains the decays of
Higgs particles into the NGBs of the aforementioned models to be completely
unobservable at colliders.  Recently, a Majoron model has been considered where
lepton number is spontaneously broken at the electroweak scale but in
accordance with astrophysical bounds \cite{pilaftsisMaj}, however we will not
consider this class of models here.  A different situation arises if we assume
that such additional NGBs, if existent, must {\it exclusively} couple to
phantom (SM gauge singlet sector) fields.

It is important to note that the requirement of renormalizability poses some 
constraints on such a hypothetical phantom sector. In particular, it demands that 
the only places where a phantom sector can make connections to the SM are
the Yukawa interactions of neutrinos and Higgs bosons and the $H^\dagger H$
``mass'' term.  Therefore, the only relevant phantom sector fields are a
right-handed fermion (possibly coming in three copies) and (in general
complex) scalar fields.  This immediately triggers some thought on implications 
for neutrino masses.  For them, there are two possibilities: Majorana or 
Dirac masses.  The Majorana see-saw mechanism \cite{seesaw} in fact is
nothing but a type of phantom sector.  However, as already discussed, in the
simplest models the possible spontaneously broken global symmetry is lepton
number -- clearly not a purely phantom sector symmetry.  So, what about the
Dirac case?  Sticking to the same principle that leads to suppressed neutrino
masses in the Majorana see-saw scenario, an analogous non-renormalizable
operator can be constructed.  It reads
\begin{eqnarray}
\mathcal{L}_{\nu} \ = \ \frac{(\overline{L}\cdot \tilde{H}) \: 
(\Phi \cdot \nu_{R})}{\Lambda}\;.
\label{eq2}
\end{eqnarray}
In the model proposed in this article, some (purely phantom sector) symmetry 
$G_{P}$, prevents the interaction $\overline{L}\cdot \tilde{H}\nu_{R}$ from 
providing neutrinos with electroweak-scale masses.  Then, \eq{eq2} results in 
acceptably small Dirac neutrino masses after spontaneous symmetry breaking of 
$G_{\rm P}$ (and the extended SM gauge group $G_{\rm SM}$) at 
$\langle \Phi \rangle \approx \langle \tilde{H} \rangle\approx 100~\mathrm{GeV}$ 
provided that $\Lambda \sim 10^{16}$ GeV.  Here, the field $H$ (where 
$\tilde{H}=i \sigma_2 H^*$) is the standard model $SU(2)_L$ Higgs
doublet and ``$\cdot$'' denotes the inner product within $G_{\rm SM}$ or
$G_{\rm P}$.  A renormalizable model resulting in the effective operator of
\eq{eq2} was first built by Roncadelli and Wyler~\cite{roncadelliwyler}. It
has been recently shown in \Ref{CDU} that this model would lead to successful
baryogenesis via Dirac leptogenesis \cite{Lindner,DiracSS,CDU} if 
$0.1~\mathrm{GeV}\lesssim \langle \Phi \rangle \lesssim 2$ TeV.

It is worth noting that this particular NGB evades many bounds applying to
other species of NGB since the only fermions transforming under $G_{\rm P}$ are the 
$\nu_{R}$, and the coupling between the NGB and the neutrinos is proportional to
$\frac{m_{\nu}}{\langle \Phi \rangle}\approx \frac{\langle \tilde{H}\rangle}{\Lambda}$. 
This is too small to affect neutrino flavour oscillations through 
$\nu \rightarrow \nu + \mathcal{J}$~\cite{MohPal}.

It is not unreasonable to suppose that the effects of the phantom sector may
already have been seen~\footnote{
	One should also notice that, like $\mathcal{J}$s, the three right-handed 
	neutrinos being SM-gauge singlets are the only light fermions that obey the 
	shift invariance, $\nu_{R}\rightarrow \nu_{R} + \omega$ where $\omega$ is 
	a Grasmann-type parameter.  It may be tempting to interpret the $\nu_{R}$s as 
	Goldstinos of an $N_{f}$ (with $n_{f}$ being the number of $\nu_{R}$ 
	flavours) supersymmetric phantom sector where the $\mathcal{J}$s belong to 
	the same supermultiplet. } 
in experiments revealing that neutrinos have small masses.

The existence of such a phantom sector may also be responsible for electroweak
symmetry breaking.  This has recently been emphasized by Patt and Wilczek~\cite{wilczek} 
and also by the authors of \Ref{CDU}.  Their argument is based on the fact that no 
symmetry principle can forbid the mixing of the Higgs sector with the phantom sector 
through the renormalizable link operator
\begin{eqnarray}
\mathcal{L}_{\rm link} \ = \ \eta \: H^{\dagger} H \Phi^{\dagger} \Phi \;.
\label{eq3}
\end{eqnarray}
\Eq{eq3} suggests that the phantom sector field $\Phi$ triggers spontaneous 
electroweak symmetry breaking, i.e.\ $\langle H \rangle \equiv v \approx 246$ GeV
once it develops a vacuum expectation value (vev), 
$\langle \Phi \rangle \equiv \sigma$.  This holds true even in the absence of any 
tree-level Higgs mass term, $\mu^{2} H^{\dagger} H$~\cite{CW,quirosespinosa,scaleinv}.  
Furthermore, it is exactly the mixing term of \eq{eq3} that causes the Higgs boson 
to decay into a pair of NGBs, $H\rightarrow \mathcal{J}\mathcal{J}$.  Since the 
$\mathcal{J}$s interact only very weakly with matter, this decay effectively 
constitutes an invisible decay of the Higgs boson.

Of course, this discussion could be generalized to non-Abelian groups.  However, 
for simplicity here and onwards the simplest group $G_{\rm P}=U(1)_{\rm P}$ is 
assumed.  The Noether current associated with this symmetry is 
$j_{\mu}(x) = i\Phi^{*}\overleftrightarrow{\partial_{\mu}}\Phi$.  
The phantom field $\Phi$ can be expanded about its vev $\sigma$ in the usual 
fashion,
\begin{eqnarray}
\Phi(x) \ = \ e^{i \mathcal{J}(x)/\sigma} \: [\sigma + \phi(x) ] /\sqrt{2} \;.
\label{eq4}
\end{eqnarray}
Using \eq{eq1} the interaction between the massive Higgs boson $\phi(x)$ and the 
NGB is found to be 
$\mathcal{L}_{\rm int}=\frac{1}{\sigma} \phi(\partial_{\mu}\mathcal{J})^{2}$.  
The scalar potential is composed of the usual quadratic and quartic terms for 
$H$ and $\Phi$ as well as the link term of \eq{eq3}. It is independent of 
$\mathcal{J}$ i.e. $V(H,\Phi) = V(h,\phi)$, where $h$ is the neutral field 
component of the $SU(2)_{L}$-Higgs doublet.  The fields $h=O_{i1} H_{i}$ and 
$\phi=O_{i2}H_{i}$ are rotated to their physical mass eigenstates, $H_{i}$, with 
an orthogonal rotation matrix $O$ (CP-conservation is assumed).  After setting 
particles on their mass shell, ${\cal L}_{\rm int}$ becomes~\cite{SS},
\begin{eqnarray}
\mathcal{L}_{\rm int} \ = \ - \frac{m_{H_{i}}^{2}}{2 \,\sigma} \: O_{i2}\: H_{i}(x) \:
\mathcal{J}(x) \: \mathcal{J}(x) \;,
\label{eq5}
\end{eqnarray}
where $i=1,2$ in this minimal $G_{\rm P}=U(1)_{\rm P}$ scenario\footnote{
	The link term of \eq{eq3} also gives rise to quartic 
	$H_{i}H_{j} \mathcal{J} \mathcal{J} (i,j=1,2)$ couplings which are
	given in Fig.\ \ref{quadverts} of~\ref{app:A}.  These couplings 
	contribute to the decay, $H_{2} \to H_{1} \mathcal{J} \mathcal{J}$.  
	However, the decay rate for this channel is on the order of $10^{-9}$ 
	GeV or less for benchmark scenarios considered in this paper.  Hence, 
	they will be completely neglected in the analysis presented here.}. 
In this case the rank-2 matrix $O$ contains one mixing angle $\theta$. Following the
notation of \Ref{CDU}, it will be fixed by $O_{12} = -O_{21} = \sin\theta$ and
$O_{11}=O_{22}=\cos\theta$.  The limit $\eta =0$ implies $\theta = 0$, i.e.\ no 
mixing between the SM-Higgs and the phantom sector scalar fields.  Obviously, in
this limit the Standard Model is recovered.  This article assumes a convention
where $m_{H1} < m_{H2}$.  Trading the vev of $\langle \Phi \rangle \equiv \sigma$ 
with the more familiar $\tan\beta \equiv v/\sigma$, the free parameters of the model
read
\begin{eqnarray}
m_{H_{1}} \quad ,\quad m_{H_{2}} \quad ,\quad \tan\theta \quad ,\quad \tan\beta \;.
\label{eq6}
\end{eqnarray}

\Eq{eq5} is the equation underlying all phenomenological analyses in this
paper.  It describes Higgs boson decays to the almost sterile NGB particles.  
There exists an extensive body of literature, which addresses various techniques 
for discovering an invisible Higgs boson at colliders.  They can be assembled in 
three main strategies:
\begin{itemize}
\item Studying the recoil of the $Z$-gauge boson in the associated $Z+H_{i}$ 
	production process. Experimental results from LEP are summarized in 
	\cite{LEP} and simulations have been performed in \cite{LEPsim}.  A study 
	for this process at the Tevatron has been performed in \Refs{MW,Han} with
	the result that the collider needs substantially more integrated luminosity 
	to improve the current LEP exclusion limit.  Parton level simulation 
	studies for the LHC exist in Refs.\cite{Kane,Godbole,Han}.  Further hadron 
	level/detector simulation studies for the LHC are currently under 
	way~\cite{Freiburg,HZinv}.
\item Vector boson fusion (VBF) processes. As suggested by Eboli and
  	Zeppenfeld~\cite{EZ}, this has now been simulated at hadron/detector level
  	for the LHC~\cite{Freiburg,VBFinv}.
\item Central exclusive diffractive production has been studied for a
	particular model in \Ref{Khoze}.
\end{itemize}
It should be noted that in all the above analyses only models with {\em only one}
Higgs boson decaying completely invisibly were considered.

In this article the focus will be put on the first two search channels, namely 
$ZH$ production and VBF.  In both cases, the coupling of the Higgs to the gauge 
bosons is crucial.  In the model considered here, only the SM-like scalar field $h$, 
belonging to the $SU(2)_{L}$ Higgs-doublet, couples to vector bosons $V$.  The
corresponding SM coupling constant  $[g_{HVV} ]_{\rm SM}$ is rescaled with the 
mixing angle such that
\begin{eqnarray}
g_{H_{i}VV} \ = \ O_{i1} \: [g_{HVV} ]_{\rm SM} \;.
\label{eq7}
\end{eqnarray}
Since the matrix $O$ is real and orthogonal, its elements are smaller than unity.  
This immediately implies that all Higgs production cross sections and/or decay 
rates (to SM particles) in this model are suppressed relative to the SM by a 
factor $O_{i1}^{2}$.  However, because of the orthogonality condition, 
$\sum_{i} [O_{i1}]^{2}=\sum_{i}[O_{i2}]^{2}=1$.  Therefore, if for example $H_{1}$ is
invisible then the other Higgs $H_{2}$ tends to be visible and vice-versa.  Is this 
a no-lose theorem for a Higgs boson discovery in this type of model?  This is one of 
the questions to be addressed in this paper.

More specifically, in this article the following two questions will be discussed:
\begin{itemize}
\item[{\cal Q}1:] Is there any window in the parameter space (\ref{eq6}) where LEP
	failed to exclude {\em both} Higgs-bosons for $m_{H_{i}} \lesssim 114$ GeV ?
\item[{\cal Q}2:] Is there any (natural) window in the parameter space
	(\ref{eq6}) where {\em both} Higgs bosons would hide undetected at LHC?
\end{itemize}
In this context ``natural'' means that the theory has a positive definite 
potential, with perturbative (non-trivial) couplings up to a high cut-off scale 
$\Lambda\approx 10^{16}$~GeV, where the mechanism for naturally light neutrino 
masses may be expected [recall \eq{eq2}].  Therefore, we begin our analysis with 
Section~\ref{theoryconstraints} where stability and triviality bounds are analyzed
and plotted together with electroweak $\rho$-parameter constraints.  In Section~\ref{Q1LEP}, 
we answer question Q1.  We derive analytical formulae for the Higgs boson to ``visible''  
($\mathcal{R}^{2}$) and  ``invisible'' ($\mathcal{T}^{2}$) decay rates and plot 
predictions of the model against experimental LEP exclusion data for Higgs masses less 
than, approximately, 114~GeV. A possible scenario explaining the LEP Higgs boson excess 
is also discussed in this section.  In Section~\ref{Q2LHC}, we extend the region of 
validity of ($\mathcal{R}^{2}$) and ($\mathcal{T}^{2}$) to heavier Higgs boson masses,
and justify five benchmark points. Next, in subsections~\ref{ZHproduction}-\ref{VBFprod}, 
we perform a detailed Monte-Carlo simulation for signals at these points and their
backgrounds, and we discuss possible strategies useful for further theoretical and 
experimental consideration.  Furthermore, in Section~\ref{sec:nona}, extensions of the Abelian  
to non-Abelian phantom sectors and some consequences relevant for Higgs boson phenomenology 
at the LHC are discussed.  A discussion of our findings together with some remarks for 
alternative scenarios is presented in Section~\ref{sec:conclusions}.  In \ref{app:A}, we 
display the relevant Feynman rules of the Abelian model.

\setcounter{equation}{0}
\section{Stability and Triviality Bounds}
\label{theoryconstraints}

In the minimal phantom model, the set of physical parameters in \eq{eq6} can
be written\footnote{We adopt the notation of \Ref{CDU}.} in terms of the renormalization 
group running parameters $\lambda_{\rm H} \, H^{4}$, $\eta\, H^{2}\: \Phi^{2}$, and 
$\lambda_{\rm \Phi}\, \Phi^{4}$:
\begin{eqnarray}
\lambda_{\rm H} &=& \frac{1}{2}\, \frac{m_{H_{1}}^{2} \cos^{2}\theta + m_{H_{2}}^{2}
\sin^{2}\theta}{v^{2}} \;, \label{eqlh} \\[3mm]
\eta &=& \frac{1}{2} \: \frac{(m_{H_{2}}^{2} -
m_{H_{1}}^{2}) \sin(2\theta) \tan\beta}{v^{2}} \;, \label{eqeta} \\[3mm] 
\lambda_{\rm \Phi} \ &=& \ \frac{1}{2}\, \frac{m_{H_{1}}^{2} \sin^{2}\theta +
m_{H_{2}}^{2} \cos^{2}\theta}{v^{2}} \; \tan^{2}\beta  \label{eqlphi}
\end{eqnarray}
with $v \approx 246$ GeV.  Notice that in the limit where both 
$\tan\beta,\, \tan\theta \rightarrow 0$ the phantom sector completely decouples 
from the SM scalar sector.   Also, note that $\lambda_{\Phi}$ depends quadratically 
on $\tan\beta$ and the Higgs boson masses.  This implies that in the case of
non-zero Higgs mixing there is always an upper bound on $\tan\beta$ if the theory 
is required to remain perturbative.  For example, if $\tan\theta =1$ and 
$m_{H} \lesssim 200$~GeV then $\tan\beta \lesssim 2$.  In all of our plots
only the case $\tan\beta =1$ is considered although, as already explained, higher
values of $\tan\beta$ would further reduce the number of visible Higgs events.

There are two\footnote{The unitarity constraint here is avoided by assuming
  that all quartic couplings are in a perturbative region, $\lambda \lesssim
  1$.}  classic, ``theoretical'' constraints on models that have been worked
out numerous times in great detail for the SM and in many of its
extensions~\cite{bounds}.  Firstly, the triviality constraint is essentially
the requirement that the couplings in \eq{eqlh} - \eq{eqlphi} stay
perturbative up to a certain scale $\Lambda_{T} \gg v$.  Secondly, the vacuum
stability constraint demands that the potential is bound from below up to a
scale $\Lambda_{V} \gg v$.  Applying both constraints yields
$\Lambda_{T},\Lambda_{V}\lesssim 10^{16}$~GeV, where we recall the discussion
following \eq{eq2}.  The vacuum stability bound can be reduced to the
requirement
\begin{equation}
4\, \lambda_H(\mathcal{Q}) \lambda_{\Phi}(\mathcal{Q}) > \eta(\mathcal{Q})^2,
\label{pos}
\end{equation}
at all scales $\mathcal{Q} \lesssim \Lambda_{V}$.
%


The running parameters are defined at the scale $\mathcal{Q}_{0}=M_{Z}$ and then 
evolved up to higher scales with the following 1-loop renormalization group 
equations~\cite{MV,Wells2}
\begin{eqnarray}
16 \pi^2\,\frac{d \lambda_H}{dt} & = & \eta^2 + 24\,\lambda_H^2 + 12\,\lambda
\,Y_t^2 - 6\,Y_t^4 
- 3\,\lambda (3\,g_2^2 + g'^2)
 +  \frac{3}{8} \Big [
2\,g_2^4 + (g_2^2 + g'^2)^2 \Big ]\,,\nonumber\\[3mm]
16 \pi^2\,\frac{d \eta}{dt} & = & \eta \Big[ 12\, \lambda_H + 8\,
\lambda_{\Phi} - 4\,\eta + 6\,Y_t 
- \frac{3}{2} (3\,g_2^2 + g'^2)
\Big]\,,\nonumber\\[3mm]
16 \pi^2\,\frac{d \lambda_{\Phi}}{dt} & = & 2\,\eta^2 + 20\,\lambda_{\Phi}^2\,.
\label{RGE}
\end{eqnarray}
Here, $t \equiv \ln \mathcal{Q}/\mathcal{Q}_0$, $g'$ and $g_2$ are the $U(1)_Y$ and 
$SU(2)_L$ gauge couplings, respectively, and $Y_t$ is the top quark Yukawa coupling.
We ignore all other Yukawa couplings because their effect in the running is negligible.
The equations for $Y_t$, $g'$ and $g_2$ are well known \cite{RGs} and are left out 
for brevity.  It is worth noticing that the parameter $\eta$ is multiplicatively 
renormalized at one loop.  Although there is no particular reason for $\eta =0$, if 
this is the case at one energy scale then this will remain true at all energy scales.

\begin{figure}[t]
 \centering
   \includegraphics[width=4.7in]{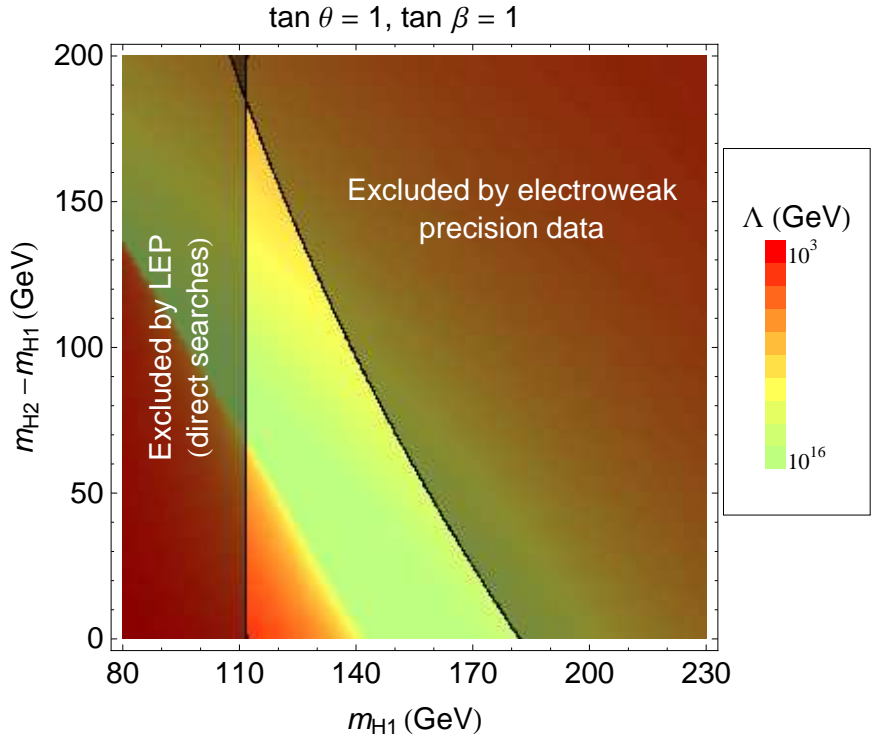}
   \parbox{0.9\textwidth}{
	\caption{The light Higgs boson mass  $m_{H_{1}}$ vs.\ $m_{H_{2}}-m_{H_{1}}$ 
	plane for $\tan \beta = \tan \theta = 1$, showing the expected cut-off 
	$\Lambda$ of the effective theory taking the triviality and positivity of the 
	potential into account (the lower of either $\Lambda_{T}$ or $\Lambda_{V}$ 
	is shown). The curved line shows the 95\% C.L.\ upper limit on the Higgs 
	boson masses stemming from precision electroweak data.}  \label{fig:lambda}}
\end{figure}

Fig.\ \ref{fig:lambda} shows the light Higgs boson mass $m_{H1}$ vs.\ $m_{H2}- m_{H1}$ 
plane for $\tan\beta = \tan \theta = 1$ where the background colours show the scale of 
new physics $\Lambda$ required either by positivity or triviality (whichever is lower).  
The curved contour shows the 95\% C.L.\ upper limit on the combined Higgs boson masses 
from precision electroweak data (see corresponding formula in \Ref{CDU}).  
Fig.\ \ref{fig:lambda} should be compared with Fig.\ \ref{fig2} of Section~\ref{Q2LHC},
to see the correspondence between easily accessible regions at the LHC and regions 
with a potentially high effective theory cut-off.  The light (light green) shaded 
parameter region of Fig.\ \ref{fig:lambda} is what we will coin the {\em natural region} 
throughout this paper. 

\setcounter{equation}{0}
\section{LEP searches}
\label{Q1LEP}
The LEP experiments searched first for visible Higgs boson events in the Higgsstrahlung 
process $e^{+} e^{-}\rightarrow Z H$ with the Higgs boson decaying to $b$-quarks and 
leptons $\ell$ in final state stemming from the $Z$ boson decay.  They 
presented~\cite{LEPres} 95\% C.L.\ upper limits for a parameter ${\cal R}^{2}$ 
(in their notation $\mbox{\rm S}_{95}$), defined as the ratio of the number of Higgs boson 
events expected in any given model to the number expected in the Standard Model 
for a Higgs boson with an identical mass, as a function of the Higgs boson mass.  An important 
point to note in this context is that $\mathcal{R}^{2}$ {\em only} counts ``visible'' 
events.  In particular, the data on decays to $b$-quarks will be used in the following.  
Then the ${\cal R}^{2}$ parameter translates into
\begin{equation}
{\cal R}^{2}_i \equiv
 \frac{\sigma(e^{+}e^{-}\to H_i\,X)\,{\rm Br}(H_i \to YY)}{\sigma(e^{+}e^{-}\to h 
\,X)\,{\rm Br}(h \to YY)}\,, \label{eqr}
\end{equation}
where $i=1,2$, $X$ are the remnants associated with the production of a $H_i$
or $h$ (the SM Higgs boson) and $YY$ could be in principle either $b\bar{b}$, or
$\tau\tau$, but {\it not} $\mathcal{JJ}$.  However, in the framework of the particular 
model studied here, another possibility is that $YY = H_{j} H_{j}$.  Exclusion 
limits in this case have been presented in~\cite{LEPres}.

The four LEP experiments~\cite{LEPin} also performed searches for acoplanar jets 
(as signal for $Z(\to q\bar{q}) \: H(\to\mathrm{invisible}$) or leptons 
(as signal for $Z(\to \ell\ell ) \: H(\to\mathrm{invisible}$), with $\ell = e,\mu$, 
apart from the DELPHI-collaboration which also used $\tau$'s in the final state.  
In all cases, the emergence of invisible decay products of the Higgs boson is 
identified with the production of missing energy ($\Emiss$).  Their study resulted 
in an upper limit on the branching ratio of $H\to \mathrm{invisible}$ as a function 
of the Higgs mass, multiplied by the production cross-section normalized to the rate 
expected from a SM Higgs decaying completely invisibly. In our case, this limit places
constraints on the parameter
\begin{eqnarray}\label{eqi}
\mathcal{T}_{i}^{2} \ \equiv \ 
\frac{\sigma(e^{+}e^{-} \rightarrow H_{i} \, X)}{\sigma(e^{+}e^{-} 
\rightarrow h \, X)} \: \mathrm{Br}(H_{i}\rightarrow \mathcal{J}\mathcal{J}) \;,
\end{eqnarray}
where again $i=1,2$, $h$ is the SM Higgs boson and $X$ are the remnants
associated with the production of $H_{i}$ or $h$ at LEP.

A further important constraint comes from the OPAL collaboration who performed
a model-\-independent analysis of the Higgs sector at LEP \cite{opalindep}. They
searched for the generic process $e^{+} e^{-} \to Z S^0$ where $S^{0}$ is a
completely neutral (and hence invisible) scalar boson.  Since this analysis is 
independent of the eventual fate of the Higgs candidate it bounds the parameter
\begin{equation}
s_i^2 \equiv \frac{\sigma (e^+ e^- \to Z H_i)}{\sigma (e^+ e^- \to Z h)}\,,
\end{equation}
as a function of the Higgs boson mass. In this model $s_1 = \cos^2 \theta$ and
$s_2 = \sin^2 \theta$.

Particularly simple expressions may be derived for $\mathcal{R}^2_{i}$ and
$\mathcal{T}^2_{i}$ in the minimal phantom scenario provided that the narrow width 
approximation may be assumed and that the Higgs boson to off-shell gauge boson decay 
modes may be neglected.  Our analytical findings closely follow the model-\-independent
analysis of \Ref{Wells1}.  Consider the case where $YY = b\bar{b}$ in \eq{eqr}.  
For simplicity let us assume that the decay $H_{2}\to H_{1}H_{1}$ is kinematically 
forbidden, i.e.\ $m_{H_{1}} > m_{H_{2}}/2$. In this case
$\mathrm{Br}(H_{i} \to b\bar{b}) +\mathrm{Br}(H_{i} \to \mathcal{JJ})\approx 1$.  
Applying this to \eq{eqr} in the LEP search region, 
$m_{H_{2}}/2 < m_{H1}\lesssim 115~\mathrm{GeV}$ and after some algebra we arrive at
\begin{eqnarray}
{\cal R}^{2}_1 & \simeq & \Bigg[(1 + \tan^2 \theta)\Big(1 +
\frac{1}{12}\,\frac{m_{H_{1}}^2}{m_b^2}\,\tan^2 \theta\,\tan^2 \beta\Big)\Bigg]^{-1}
 \;, \label{eqr1} \nonumber \\[3mm] 
{\cal R}^{2}_2 & \simeq & \Bigg[(1 + \cot^2 \theta)\Big(1 +
  \frac{1}{12}\,\frac{m_{H_{2}}^2}{m_b^2}\,\cot^2 \theta\,\tan^2 \beta\Big)\Bigg]^{-1}
  \;. \nonumber  \label{eqr2}
\end{eqnarray}
Firstly, notice that the number of Higgs boson events where the Higgs boson decays to $b\bar{b}$ 
(or indeed any other visible mode) are always suppressed relative to the SM prediction 
in which $\mathrm{Br}(h \to b\bar{b}) \approx 1$ for this particular Higgs boson mass 
region.  Secondly, the number of visible Higgs boson events decreases in the limit 
$m_{H_{i}} \gg m_{b}$. Note also that if $\tan \beta > 1$, $\mathcal{R}^2_{i}$ 
receives an additional suppression.

The importance of the Higgs boson to invisible decay and of model-\-independent Higgs 
boson analyses are highlighted when we consider the example $\tan\theta = 1$ and
$\tan\beta=2$ where we obtain $\mathcal{R}^2_{i} = 0.012$ for $m_{H_{i}}=50$ GeV.  
In principle, this is within the region allowed by LEP ``visible'' Higgs search 
data \cite{LEPres} which excludes $0.015 \lesssim \mathcal{R}_{i}^{2}\lesssim 0.2$ 
for Higgs masses in the range 
$12~\mathrm{GeV} \lesssim m_{H_{i}} \lesssim 100~\mathrm{GeV}$.

This could have been a ``nightmare'' scenario; LEP would have completely missed the 
Higgs sector!  Fortunately, this nightmare is averted by both the LEP Higgs boson to 
invisible searches and the OPAL model-\-independent Higgs boson search, because the former,
for instance, sets bounds on $\mathcal{T}_{i}^{2}$.  In the relevant LEP mass region, 
$m_{H_{2}}/2 < m_{H1}\lesssim 115~\mathrm{GeV}$, 
\begin{eqnarray}
\mathcal{T}_{1}^{2} \ &=& \ \cos^{2}\theta  \ -  \ R_{1}^{2}  \;, \label{RTtheta1}\\
\mathcal{T}_{2}^{2} \ &=& \ \sin^{2}\theta  \ -  \ R_{2}^{2}  \;.
\label{RTtheta2}
\end{eqnarray}
Setting $R^{2}_{i} \rightarrow 0$ implies that 
$\mathcal{T}_{1}^{2} + \mathcal{T}_{2}^{2} \approx 1$.  LEP searches for invisible
Higgs bosons exclude $\mathcal{T}_{i}^{2} \gtrsim 0.5$ for masses below 110 GeV,
$m_{H_{i}} \lesssim 110$~GeV.  Therefore, {\em it is unlikely that there are two 
invisible Higgs bosons in the LEP search region with masses 
$m_{H_{i}} \lesssim 110$}~GeV.  This answers question {\cal Q}1 posed in the
introduction.

In addition, using the model-\-independent analysis of OPAL \cite{opalindep},
$m_{H_i}\lesssim 85$~GeV is excluded for $s_i^2 > 0.5$.  Since either 
$s^2_{1} = \cos^2 \theta \ge 0.5$ or $s^2_{2} = \sin^2 \theta \ge 0.5$ for any 
given $\theta$, {\em OPAL excludes the case where both} $m_{H_1}\lesssim 85$~GeV 
{\em and} $m_{H_2}\lesssim 85$~GeV, {\em independently of how the Higgs bosons 
actually decay}.

It is interesting to note that one Higgs boson could still be hidden in the LEP search 
region even with these strong constraints, while the other Higgs boson then would
wait for its discovery in the allowed region out of reach of LEP.
%
\begin{center}
\begin{figure*}[t] 
\centering
\begin{tabular}{cc}
\includegraphics[width=0.48\textwidth]{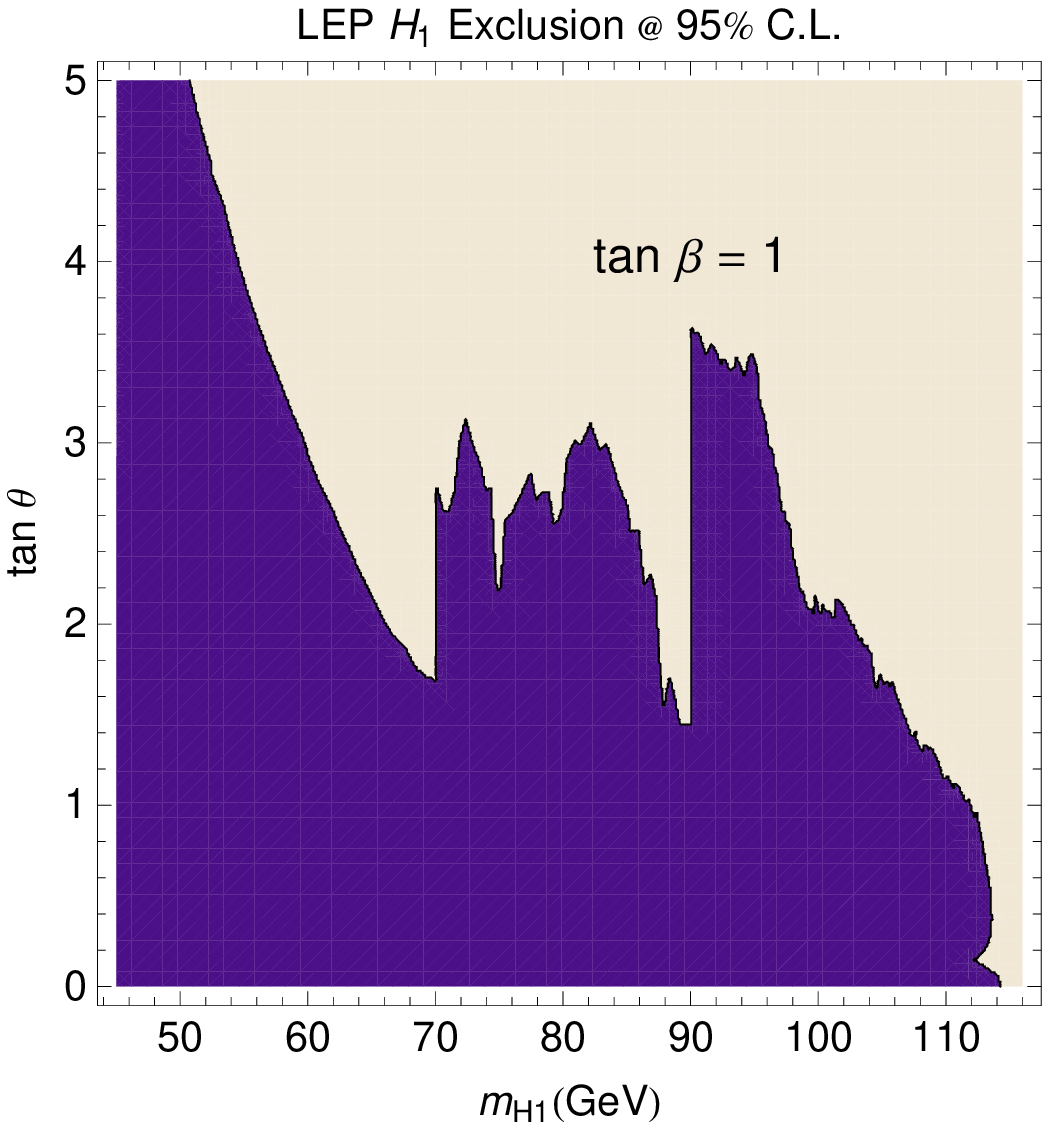} &
\includegraphics[width=0.48\textwidth]{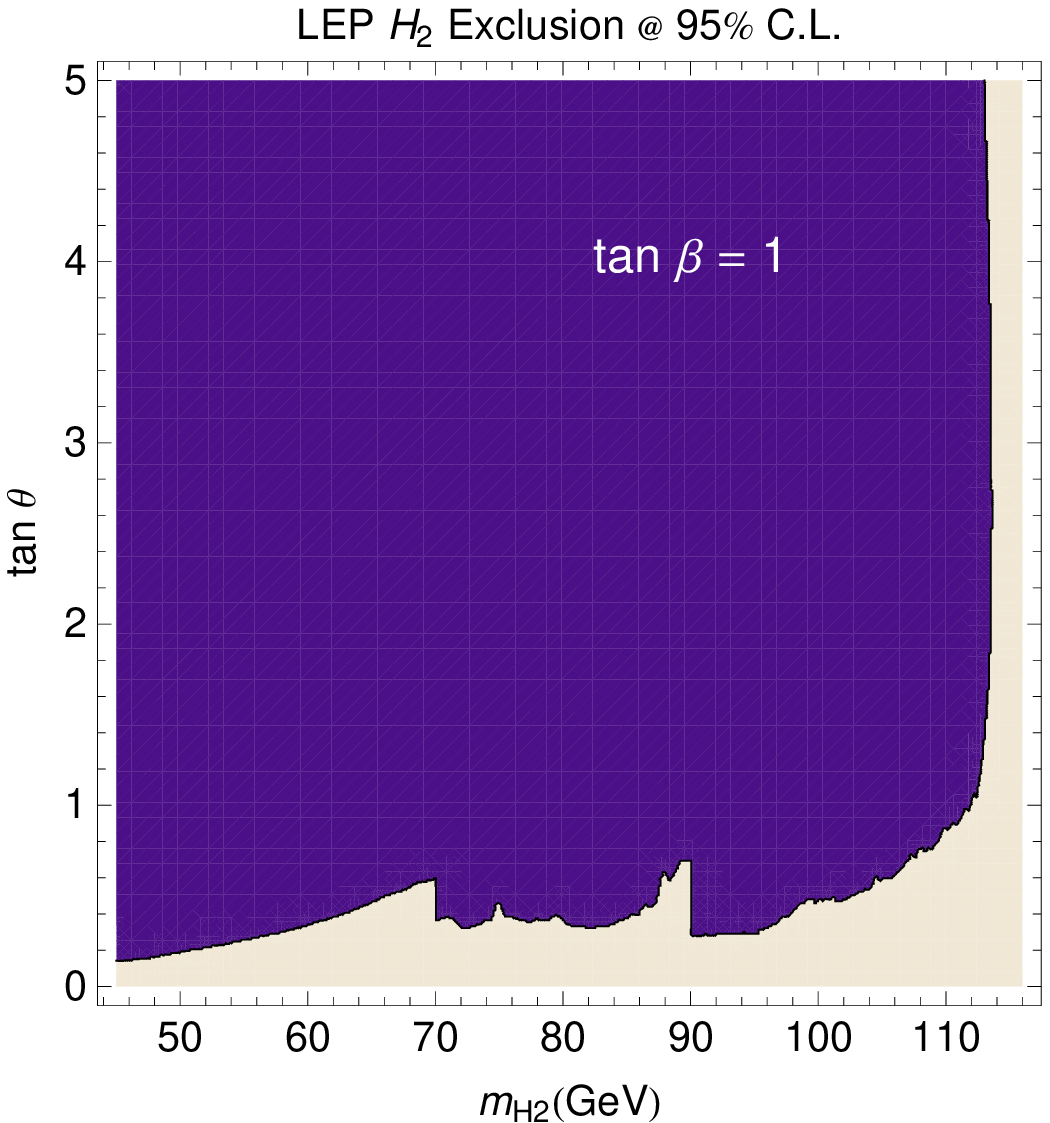} 
\end{tabular}
\parbox{0.9\textwidth}{\caption{LEP excluded regions (at 95 \% C.L.) (in dark blue) 
	for $\tan\theta$ versus Higgs boson masses $m_{H_{1}}$ (left plot) and 
	$m_{H_{2}}$ (right plot) for the minimal phantom scenario with $\tan\beta=1$. 
	{\em Both} searches are clearly complementary to each other in this scenario.}
\label{fig1}}
\end{figure*}
\end{center}
%

The results of a detailed analysis of this model, including visible, invisible and
model-independent LEP bounds~\cite{LEPres,LEPin,opalindep} are summarised in
Fig.\ \ref{fig1}. This numerical analysis confirms the analytical findings above.  
With $\tan\beta =1$, a light Higgs boson ($H_{1}$) with a mass as low as 65 GeV 
could have escaped unobserved at LEP if $\tan\theta \gtrsim 2$. For the same range 
of $\tan \theta$ the other Higgs ($H_{2}$) is constrained to be heavier than 114 
GeV.  From inspection of Fig.\ \ref{fig1} we can define a LEP-allowed benchmark 
scenario B1 for the phantom model presented here, namely:
\begin{eqnarray}
\mathrm{B1:} \quad m_{H_{1}} = 68~\mathrm{GeV} \quad &,& \quad 
m_{H_{2}} = 114~\mathrm{GeV} \quad , \quad \nonumber  \\[2mm]
\tan\theta = 2 \quad &,& \quad \tan\beta = 1 \;.
\label{lepben}
\end{eqnarray}  
In this case one Higgs boson is buried, undiscovered in the LEP search region
due to the small values of $\mathcal{R}_{1}^{2}$ and $\mathcal{T}_{1}^{2}$
which have to satisfy $\mathcal{R}_{1}^{2} + \mathcal{T}_{1}^{2} = \cos^2 \theta
= 0.2$ following eq.~(\ref{RTtheta1}). The other, heavier Higgs has
$\mathcal{R}_{2}^{2} = 0.06$ and $\mathcal{T}_{2}^{2} = 0.74$.  With this set
of parameters, very few $H_2$ events are SM-like decays into ``visible'' final
states and instead $H_2$ decays mainly into ``invisible'' NGBs. This scenario
could well be classed as a (LEP) nightmare!

As yet, no combined LEP limits exist on invisibly decaying Higgs bosons with
masses below $m_H = 90$~GeV. Therefore, for $m_H < 90$~GeV the limits
presented here are based on the individually published results from each
experiment. However, some of the individual studies do not cover the whole
Higgs mass range considered here and so the best available limit is used for
any given Higgs mass. This is one of the causes of the sharp edges in Fig.\
\ref{fig1}. Clearly, a future combined LEP analysis may well exclude the
benchmark B1 which lies close to being ruled out by ALEPH \cite{LEPin} which
considered Higgs masses down to $m_H = 70$~GeV for which $\mathcal{T}^2 \simeq
0.1$ is excluded.

\subsection{A digression: 2.3 $\sigma$ LEP Higgs search excess}

\begin{figure}[t]
\centering
\includegraphics[scale=0.83]{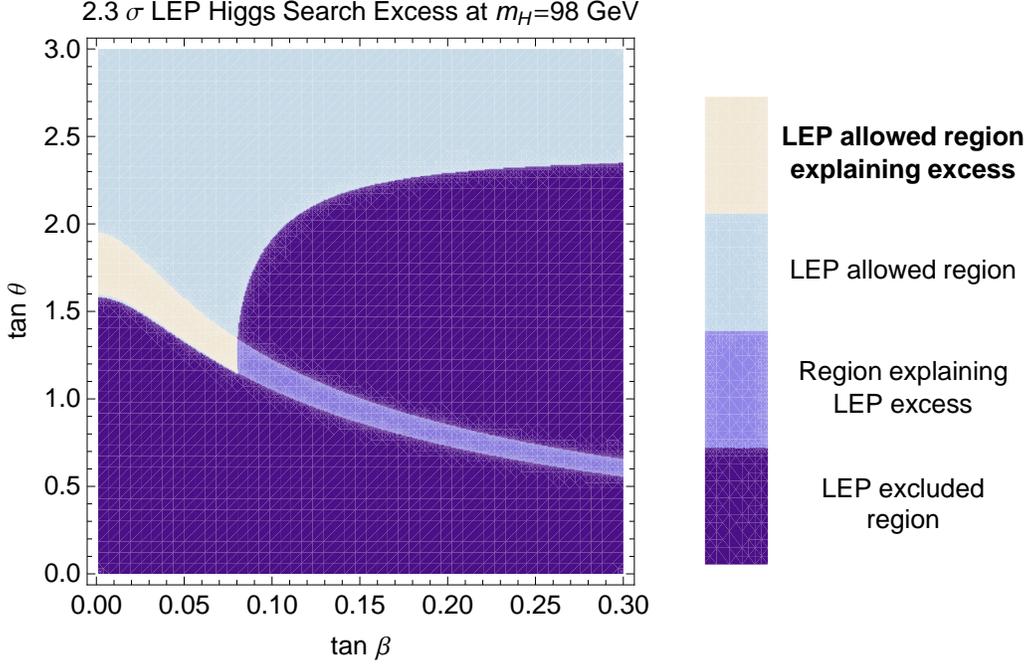}
\parbox{0.9\textwidth}{\caption{The $\tan \theta$ vs. $\tan \beta$ plane for a lightest 
	Higgs mass of $98~{\rm GeV}$.  The lightest region indicates where the 
	$2.3 \sigma$ effect in the LEP Higgs searches could be explained whilst still 
	being consistent with other LEP Higgs search data (such as the search for invisible 
	Higgs bosons).}
\label{fig:lepexcess}}
\end{figure}

The LEP experiments established a small, 2.3$\sigma$ effect in their Higgs boson searches 
corresponding to a Higgs boson mass of about 98~GeV \cite{lepexcess}.  Explaining this excess 
would require a value of ${\cal R}_1^2 \simeq 0.2$, ruling out a Standard Model Higgs 
boson as plausible explanation.  It is possible to provide a candidate Higgs boson in 
the phantom model discussed in this publication, which would have produced such an effect 
in the LEP data.  Fig.\ \ref{fig:lepexcess} shows the allowed region in the 
$\tan \theta$ vs. $\tan \beta$ plane for $m_{H_1} = 98$~GeV.  The allowed region 
is tightly constrained because of the searches for invisible Higgs bosons at LEP 
in this mass region.  At the relatively small values of $\tan \beta$ still allowed, the 
main reason for such a small value of ${\cal R}_1^2$ is Higgs mixing rather than 
the extra invisible decay mode suppressing the Higgs branching ratio.

\begin{figure}[t]
\centering
\includegraphics[scale=0.98]{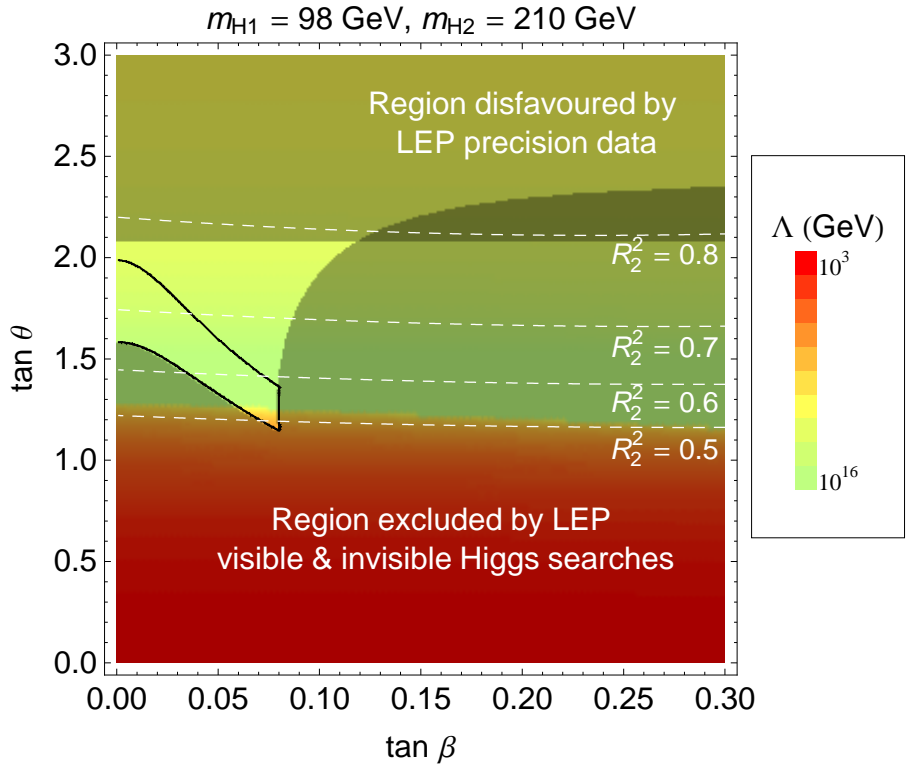}
\parbox{0.9\textwidth}{\caption{The $\tan \theta$ vs. $\tan \beta$ plane for Higgs 
	boson masses of $m_{H1} = 98~{\rm GeV}$ and $m_{H2} = 210~{\rm GeV}$.  The region 
	enclosed by the black line indicates where the $2.3 \sigma$ effect in the LEP Higgs
  	searches could be explained whilst still being consistent with other LEP data for
	Higgs boson searches.  The background colours indicate the scale of the expected
  	cut-off $\Lambda$, of the effective theory taking the triviality and
  	positivity of the potential into account.  Darkly shaded regions are excluded
  	by LEP Higgs search data. Contours for ${\cal R}^2_2$ are shown in white.}
\label{fig:lepexcesstriv}}
\end{figure}
Fig.\ \ref{fig:lepexcesstriv} shows the constraints on this region of
parameter space coming from considering the triviality and positivity of the
potential.  For a suitably heavy $m_{H2} \stackrel{>}{{}_\sim} 210$~GeV, most
of the region suggested by the LEP excess is described by a theory which could
be valid to scales as high as $10^{16}$~GeV.  When $\tan \theta \sim 1$, as
$\tan\beta \to 0$ ($\sigma \to \infty$) the values of $\eta$ and
$\lambda_\Phi$ tend to 0. Looking at eqs.~(\ref{RGE}) it is apparent that
small values of $\eta$ and $\lambda_\Phi$ will be relatively stable under
renormalization group evolution since, for example, $\eta$ is multiplicatively
renormalized. Higgs masses around the electroweak scale are maintained in this
limit because $\mu_{H}^2 \to \infty$ whilst $\mu^{2}_{\Phi} \sim -v^2$.
However, because $\eta \to 0$ the $\Phi$ and $H$ sectors are almost decoupled
so that potentially destabilizing diagrams with a heavy $H$ will be
proportional to $\eta$ and not greatly affect the mass of $\Phi$. Of course,
the model is still quadratically sensitive to a high cut-off scale and thus
still possesses the hierarchy problem of the SM.

Note that the second Higgs boson mass is restricted by the upper limit on Higgs
boson masses from precision electroweak data \cite{CDU}, however for $m_{H2}
\stackrel{<}{{}_\sim} 210$~GeV the whole region suggested by the LEP excess is
free from this constraint. Clearly further data would be required before this
effect could be taken more seriously.

In the next chapter we will address the question of whether the LHC has the
sensitivity required to discover these scenarios, in particular the potential
nightmare B1. The possible existence of other challenging  scenarios with heavier
Higgs bosons will also be examined.

\section{LHC: expectations and strategic searches}
\label{Q2LHC}

In the LHC search region, the parameters $\mathcal{R}_{i}^{2}$ and 
$\mathcal{T}_{i}^{2}$ can be defined by expressions similar to those in 
\eqs{eqr}{eqi}, respectively, with the obvious replacement of the
electron/positron initial state to a proton/proton initial state and $YY =
\gamma \gamma , b\bar{b}, VV, gg,$ etc..
Two categories for the ratios ${\cal R}_{i}^{2}$ may be distinguished: ({\it a})
a region where $m_{H_{i}}< 2\, m_{V}$ and $H_{i}$ decays dominantly into $b\bar{b}$ 
and ({\it b}) a region where $m_{H_{i}}\stackrel{>}{\sim} 2\, m_{V}$ and the 
$H_{i}$ decays dominantly into a gauge boson pair, $VV$, with $V=Z,W$.  

In case ({\it a}), and under the assumption that gauge bosons are produced on-shell, 
analytical approximations for $\mathcal{R}^{2}_{i}$ are identical to those studied 
in the previous chapter.  On the other hand, assuming a common gauge boson mass $m_{V}$, 
in region ({\it b}), we obtain
\begin{eqnarray}
{\cal R}^{2}_1 & \simeq & \Bigg[(1 + \tan^2 \theta)\Big(1 +
  \frac{1}{3\,g(x_1)}\,\tan^2 \theta\,\tan^2 \beta\Big)\Bigg]^{-1} \;, \nonumber \\[3mm]
{\cal R}^{2}_2 & \simeq & \Bigg[(1 + \cot^2 \theta)\Big(1 +
  \frac{1}{3\,g(x_2)}\,\cot^2 \theta\,\tan^2 \beta 
+ \frac{f(y)}{g(x_2)}\,\frac{\cot^2 \theta}{(1+\cot^2 \theta)^2}\,(\cot \theta
  - \tan \beta)^2\Big)\Bigg]^{-1}  \;, \nonumber \\[3mm]
  \label{lhcr}
\end{eqnarray}
where $x_i = m^2_V/m^2_{Hi}$, and $g(x) = (1 - 4 x + 12 x^2)\,(1-4x)^{1/2}$.
The last term in \eq{lhcr} is the contribution from the heavy  Higgs boson
decay $H_2 \to H_1 H_1$~\cite{Wells2,Grossman}.  Furthermore, $y = m_{H1}^2/m_{H2}^2$ 
and $f(y) =(1 + 4 y + 4 y^2)\,(1-4y)^{1/2}\,\Theta(1-4y)$.  Imposing some constraints 
to this analysis (see section \ref{theoryconstraints}), the mode $H_{2}\to H_{1} H_{1}$ 
will not be important in further discussions.

It is apparent from \eq{lhcr} that a certain suppression of the observable
rates ($\mathcal{R}^{2}_{i}$) is always present. Its origin is twofold.  Firstly, 
the couplings between the $H_{i}$ and SM fields are always suppressed because of 
the mixing matrix $O$.  Secondly, the decay widths of the Higgs bosons are 
enhanced by the extra decay mode $H_{i} \to {\cal JJ}$.  The contribution of this 
additional decay mode is increased at large $\tan\beta$ and for $\tan\beta = 10$ 
and $\tan \theta = 1$ the suppression of visible events is always more than 90\% 
for $m_{H_{2}} \lesssim 200$~GeV.  However, as we have already remarked in 
section \ref{theoryconstraints}, high values of $\tan\beta$ result in non-perturbative 
couplings and will therefore not be considered in this article.

What then would be a nightmare scenario for the LHC?  At present both the ATLAS and
CMS collaborations have performed studies, at detector simulation level, to explore 
the discovery potential of their apparatus for both SM-like Higgs bosons which decay 
to visible final states, see e.g.~\cite{Mangano}, and Higgs bosons decaying to 
invisible final states, for example~\cite{VBFinv,HZinv}.  These studies are 
sensitive to the ratios ${\cal R}_{i}^{2}$ and ${\cal T}_{i}^{2}$ as functions 
of the Higgs boson mass.  For example, looking at the simulation results for the LHC 
with $\mathcal{L}=10(30)~\mathrm{fb}^{-1}$ integrated luminosity we estimate (with 
na\"ive scaling) that it would be difficult to discover a visibly-decaying Higgs if 
signal event rates were 30\%(20\%) of that expected in the SM 
(${\cal R}_{i}^{2} \lesssim 0.3(0.2)$).  Furthermore, studies of the 
sensitivity of the ATLAS detector\footnote{Similar studies exist for the CMS 
	detector~\cite{Clare}.}  to invisibly decaying Higgs bosons suggest that 
after $\mathcal{L}=10(30)~\mathrm{fb}^{-1}$ integrated luminosity ATLAS could
exclude Higgs bosons with ${\cal T}_{i}^{2} \gtrsim 0.30(0.25)$ at 95\%
C.L.\ \cite{VBFinv,HZinv}.

\begin{figure}[t]
\centering
   \includegraphics[width=.6\textwidth]{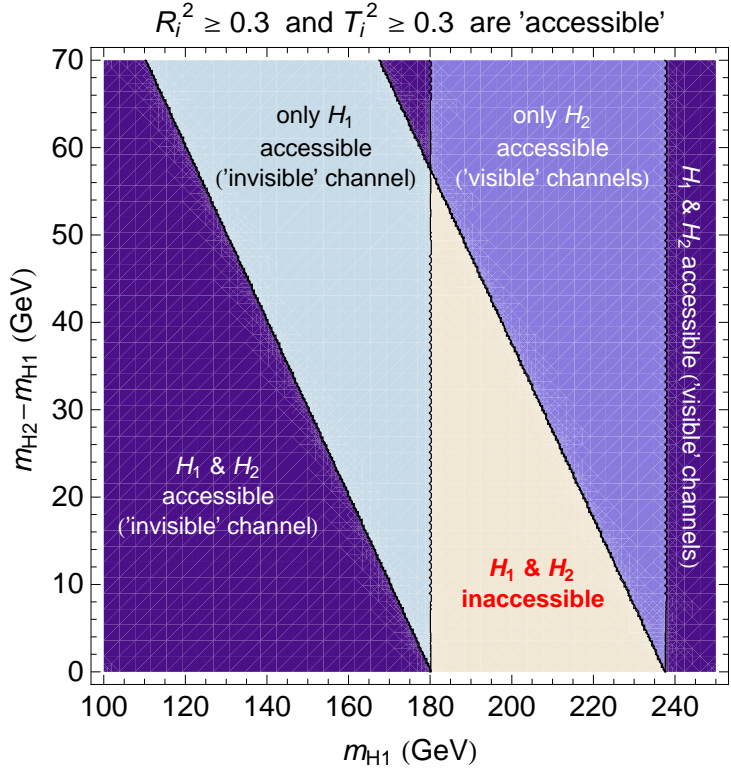} 
   \parbox{0.9\textwidth}{\caption{Regions in the $m_{H_{1}}$ vs. $m_{H_{2}}-m_{H_{1}}$ 
	plane for $\tan \beta = 1$ and $\tan \theta = 1$, where different Higgs bosons are
	``accessible'' at LHC.  We define that a given $H_{i}$ is accessible if either
	$\mathcal{R}_{i}^{2} \ge 0.3$ or $\mathcal{T}_{i}^{2} \ge 0.3$. In the
	dark (blue) regions both Higgs bosons are accessible. In the white
	(beige) region no Higgs bosons are accessible.}
  \label{fig2}}
\end{figure}

To further illustrate the necessity of the Higgs boson to invisible searches in this minimal
phantom scenario, in Fig.\ \ref{fig2} areas on the $m_{H_{1}}$ vs.\ $m_{H_{2}}
- m_{H_{1}}$ plane are plotted where $\mathcal{R}_{i}^{2}\ge 0.3 $ and/or
$\mathcal{T}_{i}^{2} \ge 0.3$.  These limits define na\"ive regions, where Higgs
bosons will experimentally be accessible at the LHC, either in visible or
invisible search channels.  Clearly at this stage in this study these limits
are assumptions, and in fact the true experimental reach of the LHC will not
be known until after it has been running for some time and predictions for the
levels of backgrounds have been confirmed (or not). These assumptions do,
however, serve as a good first estimate on which to justify the further study
undertaken here.

In producing Fig.\ \ref{fig2} all Higgs boson decay modes including decays to
off-shell vector bosons have been considered.  Different colours indicate
regions where either one, both or no Higgs bosons can be seen in different
channels.  It is clear that a truly challenging region for LHC region remains
where $\mathcal{R}_{i}^{2}\le 0.3$ and $\mathcal{T}_{i}^{2} \le 0.3$. This
motivates the further more detailed Monte Carlo analysis in the later sections
of this article, which probe more carefully the possibility of discovering a
Higgs boson in the invisible search channel when $\mathcal{T}_{i}^{2}
\stackrel{<}{{}_\sim} 0.3$.

Using equations~(\ref{RTtheta1}) and (\ref{RTtheta2}) it is easy to see that
$\mathcal{R}_1^2 + \mathcal{R}_2^2 + \mathcal{T}_1^2 + \mathcal{T}_2^2 = 1$.  The 
following no-lose theorem then exists: If experiments can discover a Higgs boson 
over the whole range of $\mathcal{R}_i^2$ down to $\mathcal{R}_i^2 = 0.25$ {\em or} 
over the whole range of $\mathcal{T}_i^2$ down to $\mathcal{T}_i^2 = 0.25$ then 
at least one Higgs boson should be found.
 
Without real data, estimates of the capabilities of experiments like ATLAS and CMS 
may easily be too optimistic or too pessimistic.  Therefore in this publication, a 
constructive approach is taken.  The phantom model has been added to the Monte Carlo 
event generator \Sherpa \cite{sherpa}, ready to be used when real data arrive. For
now Fig.\ \ref{fig2} may be used to define additional benchmark scenarios, some in
potentially nightmarish regions, and these points can be studied in more
detail. The particular scenarios are displayed in Table~\ref{tab:1}.

\begin{table*}[t]
\vspace{0.5cm}
   \centering
   \begin{tabular}{|c||c|c|} 
      \hline
      \multicolumn{3}{|c|}{$\tan\theta = 1 \quad \tan\beta =1$} \\
	\hline
      {Benchmark} & $m_{H_{1}}$(GeV) & $m_{H_{2}}$(GeV) \\
     \hline
      B2  & 112 & 130 \\
      B3  & 140 & 165 \\
      B4  & 160 & 190 \\
      B5  & 185 & 190 \\
      \hline
   \end{tabular}
   \parbox{0.9\textwidth}{\caption{Four LHC benchmark scenarios for the phantom model.}
   \label{tab:1}}
\end{table*} 
%
The LO branching ratios for both Higgs bosons are presented in Table~\ref{tab2}.  These ratios 
are in agreement with the analytical LO expressions in \eq{lhcr} and Fig.\ \ref{fig2}, and 
the discussion following them.  The most optimistic benchmark point is B1 and the most 
challenging one is B5.   	%
\begin{table*}[t]
\centering
	\begin{tabular}[h]{|c|c|
				r|r|r|r|r|}
\hline
Benchmark &  Higgs   & $\Gamma_{tot}$(GeV) &  
\parbox{15mm}{$b \bar{b}$} & 
\parbox{15mm}{$W^{+}W^{-}$} & 
\parbox{15mm}{$ZZ$} & 
\parbox{15mm}{$\mathcal{J} \mathcal{J}$} \\
\hline \hline
B1 & $H_{1}$ & 0.041 & 0.694 & --  & -- & 99.222 \\     
     & $H_{2}$ & 0.051 & 3.567 &  0.289 & 0.020 & 95.533 \\  
\hline
\hline  
B2 & $H_{1}$ & 0.117 & 0.958 &  0.059 & 0.003 & 98.823 \\
     & $H_{2}$ & 0.183 & 0.697 &  0.348 & 0.042 & 98.784 \\
\hline
\hline
B3 & $H_{1}$ & 0.229 & 0.593 &  0.779 & 0.103 & 98.408 \\
     & $H_{2}$ & 0.490 & 0.319 &   23.769 & 0.498 & 75.339 \\ 
\hline
\hline 
B4 & $H_{1}$ & 0.387 & 0.393 &  12.217 & 0.396 & 86.904 \\
   & $H_{2}$ & 1.066 & 0.166 &  36.597 & 10.313 & 52.879\\
\hline
\hline 
B5 & $H_{1}$ & 0.921 & 0.188 &  36.500 & 6.787 & 56.475 \\
   & $H_{2}$ & 1.066 & 0.166 &  36.597 & 10.31 & 52.879 \\  
\hline
\end{tabular}
\parbox{0.9\textwidth}{\caption{Branching ratios (in percent) and total widths (in units of GeV) 
for the Higgs bosons, $H_{i}(i=1,2)$, for the benchmark points of Table~\ref{tab:1}.  
Branching ratios that are not displayed, account for less than 0.4\%. }
\label{tab2}}
\end{table*}

Prospects for discovering the Higgs bosons in the various benchmark scenarios B1-B5 at the LHC 
will be studied in the following.  Theoretical vacuum stability and triviality bounds as well 
as bounds from fitting electroweak (EW) observables have already been presented in 
Section~\ref{theoryconstraints}.  All benchmark scenarios selected in Table~\ref{tab:1} satisfy 
the EW constraints and in some the effective theory may be valid even to scales as
high as the Planck scale.

\subsection{$ZH$-production}
\label{ZHproduction}

The first search channel for an invisibly decaying Higgs boson at the LHC considered
here is the associated production of a $Z$ and a Higgs boson, where the $Z$ decays 
leptonically.  This ensures that a corresponding event can be triggered.  The backgrounds 
to this process include $ZZ$, $WW$, $WZ$ and $Z$ production with corresponding decays, 
and fully leptonic $t\bar t$ production\footnote{
	Note that, in all processes, off-shell effects, $Z$-$\gamma$ interference
	etc.\ are fully included in the simulation.}.  
It should be noted here that in principle some information on the rates 
can be obtained directly from data: for $ZZ$ pairs, final states with four
leptons may be reweighted with the corresponding $Z\to\nu\bar\nu$ branching ratio,
in the $WW$ case, different sign, different lepton pairs may be invoked.  For the 
$WZ$ background, it may be possible to extrapolate from events where three leptons
are seen to those where one lepton is lost, i.e.\ either outside the detector 
acceptance or undetected.  For top-pair production, semi-leptonic events may help.  

All processes have been simulated with \Sherpa~\cite{sherpa} in the following setup: 
In order to correctly model hard parton radiation \Sherpa employs the multijet matrix 
element-parton shower merging procedure of \cite{CKKW}.  Therefore, for all processes 
discussed here and in the next section, matrix elements with at least one and in most 
cases two additional jets have been added to the simulation.  This ensures that the 
simulation correctly describes the important high-$p_\perp$ tails of various distributions.  
However, all cross sections quoted are, in principle, obtained at leading-order accuracy, 
with no $K$-factors added to them.  CTEQ6L parton distubution functions are used with 
$\alpha_{s}(M_Z)=0.118$ \cite{Pumplin:2002vw}.  $\alpha_{s}$ is computed at two--loop accuracy.  
All scales are set according to the merging prescription of \cite{CKKW}.  Jets have been 
defined in all cases through the $k_{T}$ algorithm \cite{KTjets}.  The CKM matrix has been 
choosen to be diagonal.

We have simulated and analysed events with electrons in the final state; mostly identical 
numbers would have been obtained if we had specialised for muon pairs instead.  Obviously, 
this difference would be of great importance if detector effects had been included as well\footnote{
	We refrained from including full detector simulations, or any Gaussian smearing or
	electron-jet conversion ``by hand'' and concentrated on an analysis at the hadron 
	level, including all effects of fragmentation, hadron decays, final state QED 
	bremsstrahlung etc..}.  
However it should suffice to state that we quote final results for leptons $\ell=e,\mu$. 
We also omitted all effects due to the underlying event because of the large uncertainties
related to its modelling and the rather small impact it has on the observables we discuss.

The selection cuts listed in Ref.~\cite{HZinv} have been applied.  Thus we require:
\begin{enumerate}
\item one lepton pair of the same kind with opposite charges, where each lepton individually 
	satisfies $p_{T,\ell}>15$ GeV and $|\eta_\ell|<2.5$;
\item $|M_{\ell\bar\ell}-M_Z|\le 10$ GeV;
\item $\ETmiss > 100$ GeV;
\item a veto on jets with $p_T>20$ GeV, $|\eta|<4.9$; 
\item a veto on b-jets with $p_T>15$ GeV, $|\eta|<4.9$;
\item $m_T>200$ GeV, where $m_T=\sqrt{2p_T^{\ell\bar\ell}\pTmiss (1-\cos\phi)}$.
\end{enumerate}
Additionally, we impose:
\begin{enumerate}
\item[6.] $\Delta R_{\ell\bar\ell}<1.75$;
\item[7.] $p_{T}(\ell\bar\ell\ETmiss)<60$ GeV.
\end{enumerate}

For the various backgrounds listed above, cross sections before and after these additional 
selection cuts are listed in Table~\ref{Tab:ZHbackground}.
Generation cross sections, selection cut efficiencies and the resulting selection cross 
sections for the signal in the different benchmark scenarios are given in 
Table~\ref{Tab:ZHsignal}.  It should be stressed again that all cross sections quoted have
been obtained at leading order accuracy.

\begin{table*}
\begin{center}
\begin{tabular}{|l||c|c|c|c|c|}
\hline
& \parbox{15mm}{$ZZ$} 
& \parbox{15mm}{$W^\pm Z$}
& \parbox{15mm}{$W^+W^-$}
& \parbox{15mm}{$t\bar t$}
& \parbox{15mm}{$Z$}\\\hline\hline
$\sigma_{\rm tot}^{\rm gen}$ [fb] & 164 & 1.17$\cdot 10^{3}$ & 
	1.01$\cdot 10^{4}$ & 7.44$\cdot 10^{4}$ &1.81$\cdot10^{6}$ 
\\\hline\hline
$\ell^+\ell^-$ only         & 2.00$\cdot 10^{-1}$ & 1.10$\cdot 10^{-1}$  & 
  6.59$\cdot 10^{-2}$ & 8.40$\cdot 10^{-2}$ & 1.41$\cdot 10^{-1}$ \\
$|m_{\ell \ell}-M_Z|<10$ GeV & 1.87$\cdot 10^{-1}$ & 9.17$\cdot 10^{-2}$  & 
  8.92$\cdot 10^{-3}$ & 1.09$\cdot 10^{-2}$ & 1.25$\cdot 10^{-1}$ \\
$E\!\!\!/_T>100$ GeV  & 3.69$\cdot 10^{-2}$ & 1.10$\cdot 10^{-2}$  & 
  5.91$\cdot 10^{-4}$ & 2.41$\cdot 10^{-3}$ & 1.94$\cdot 10^{-7}$ \\
jet veto              & 1.64$\cdot 10^{-2}$ & 2.13$\cdot 10^{-3}$  & 
  3.53$\cdot 10^{-5}$ & 2.00$\cdot 10^{-6}$ & -                   \\
$m_T>200$ GeV         & 1.54$\cdot 10^{-2}$ & 1.95$\cdot 10^{-3}$  & 
  2.74$\cdot 10^{-5}$ & 1.19$\cdot 10^{-9}$ & -                   \\
$\Delta R_{\ell\ell}<1.75$, 
$p_{T}(\ell\ell,\ETmiss)<60$ GeV  & 1.23$\cdot 10^{-2}$ & 1.50$\cdot 10^{-3}$  & 
  2.23$\cdot 10^{-9}$ & 1.55$\cdot 10^{-10}$ & -                   \\
\hline\hline
$\sigma_{\rm eff}$ [fb] & 2.02 & 1.75 & 2.25$\cdot 10^{-5}$ & 1.15$\cdot 10^{-5}$ & - \\\hline
\end{tabular}
\parbox{0.9\textwidth}{\caption{Generation characteristics for the background processes to 
	the $ZH$-channel.  In all cases we included all leptonic decay modes: In the $ZZ$ case, 
	therefore the final state included a lepton and a neutrino pair, in the $WZ$ case, we 
	included a lepton pair from the $Z$ and a lepton-neutrino pair from the $W$, the $WW$ 
	channel was supposed to decay fully leptonically in all possible combinations, for the 
	top pairs we assumed purely leptonic decays, and for the $Z$ a leptonic final state 
	(no neutrinos) was demanded.}
\label{Tab:ZHbackground}}
\end{center}
\end{table*}

\begin{table*}
\begin{center}
\begin{tabular}{|l||c|c|c|c|c|}
\hline
                     & $B_1$   & $B_2$   & $B_3$   & $B_4$   & $B_5$   \\\hline \hline
$\sigma_{\rm tot}$ [fb]        
                     & 280     & 114.6   & 53.0    & 29.0    & 13.6    \\\hline \hline
$\ell^{+} \ell^{-}$ only        & 1.75$\cdot 10^{-1}$ & 1.98$\cdot 10^{-1}$ & 2.25$\cdot 10^{-1}$ & 2.40$\cdot 10^{-1}$ & 2.24$\cdot 10^{-1}$ \\
$|m_{\ell \ell}-M_Z|<10$ GeV & 1.62$\cdot 10^{-1}$ & 1.84$\cdot 10^{-1}$ & 2.10$\cdot 10^{-1}$ & 2.23$\cdot 10^{-1}$ & 2.08$\cdot 10^{-1}$ \\
$E\!\!\!/_T>100$ GeV & 3.12$\cdot 10^{-2}$ & 6.07$\cdot 10^{-2}$ & 8.91$\cdot 10^{-2}$ & 1.08$\cdot 10^{-1}$ & 1.12$\cdot 10^{-1}$ \\
jet veto             & 3.00$\cdot 10^{-2}$ & 5.66$\cdot 10^{-2}$ & 7.85$\cdot 10^{-2}$ & 9.29$\cdot 10^{-2}$ & 1.08$\cdot 10^{-1}$ \\
$m_T>200$ GeV        & 2.88$\cdot 10^{-2}$ & 5.49$\cdot 10^{-2}$ & 7.64$\cdot 10^{-2}$ & 9.08$\cdot 10^{-2}$ & 1.06$\cdot 10^{-1}$ \\
$\Delta R_{\ell\ell}<1.75$, 
$p_{T}(\ell\ell,\ETmiss)< 60$ GeV  & 2.55$\cdot 10^{-2}$ & 4.93$\cdot 10^{-2}$ & 6.94$\cdot 10^{-2}$ & 8.35$\cdot 10^{-2}$ & 9.85$\cdot 10^{-2}$ \\
\hline\hline
$\sigma_{\rm eff}$ [fb] 
                     & 7.15    & 5.65   & 3.68    & 2.42   & 1.34    \\\hline
\end{tabular}
\parbox{0.9\textwidth}{\caption{
Generation characteristics for the signal processes in the $ZH$-channel. In each case we
assumed all leptonic decay channels for the $Z$ boson.}
\label{Tab:ZHsignal}}
\end{center}
\end{table*}

The numbers from both Tables \ref{Tab:ZHbackground} and \ref{Tab:ZHsignal} suggest that
the two most dangerous backgrounds to the $ZH$ signal are $ZZ$ and $WZ$ production, with
corresponding decays.  Following our discussion above, however, it seems that the total 
cross sections and distributions related to these backgrounds can be directly extracted 
from data in the $ZZ$ case or probably well extrapolated from measurements.  After cuts 
we find that the backgrounds together account for roughly 8 fb, leaving us with 
signal-to-background ratios of the order of $S/B\approx 1/8$ up to 1.  We therefore 
conclude that it should be possible to find the signal in all five benchmark scenarios.
However, we would like to stress here that more conclusive numbers can be obtained
after a simulation at detector level only.

Such detector-level studies for an invisibly decaying Higgs boson have been
for the ATLAS experiment \cite{HZinv} found signal--\-to--\-background ratios reaching up
to $1/4$.  Although this is of the same order of magnitude as our results, there
are several differences: First of all, in our simulation the \Sherpa Monte Carlo event
generator with multijet merging was used for both signal and background events, while
the ATLAS study employed the \Pythia \cite{Pythia} event generator for the backgrounds
and the program \htohv~\cite{h2hv} for the signal.  While \Sherpa and \Pythia are
formally of the same accuracy there are a number of differences, like \Sherpa multijet 
merging leading to an improved treatment of hard QCD radiation, and the full inclusion
of spin correlations in \Sherpa, which are not present in \Pythia.  This may have lead 
to a better separation of signal and background in \Sherpa.  On the other hand, in
ATLAS' simulation the $HVV$ couplings where assumed to have exactly the same strength
as in the SM - which is not true for our analysis, where these couplings are reduced due 
to mixing effects.  In addition, a 100\% branching ratio of Higgs boson to invisible
was assumed for the ATLAS simulation, again in contrast with our simulation, where
the relevant branching ratio ranged between roughly 50\% up to 100\%.  These two facets
of the study, of course, enhance the signal--\-to-\-background ratio in the ATLAS study.
Of course, there are further differences, like the missing underlying event in \Sherpa,
which has been included in the ATLAS study, like slightly different selection cuts, like
a different choice of PDF (CTEQ5L in ATLAS, CTEQ6L in our study) and, most importantly, 
like the inclusion of detector effects through their fast detector simulation ATLFAST 
\cite{atlfast} in the ATLAS study that are totally absent in our case.  To summarize: 
However different in detail the studies are, it is reassuring to see that in all cases 
this seems to be a feasible channel, at least at accumulated higher luminosities.

\begin{figure}[t]
\centering
\includegraphics[width=0.49\textwidth]{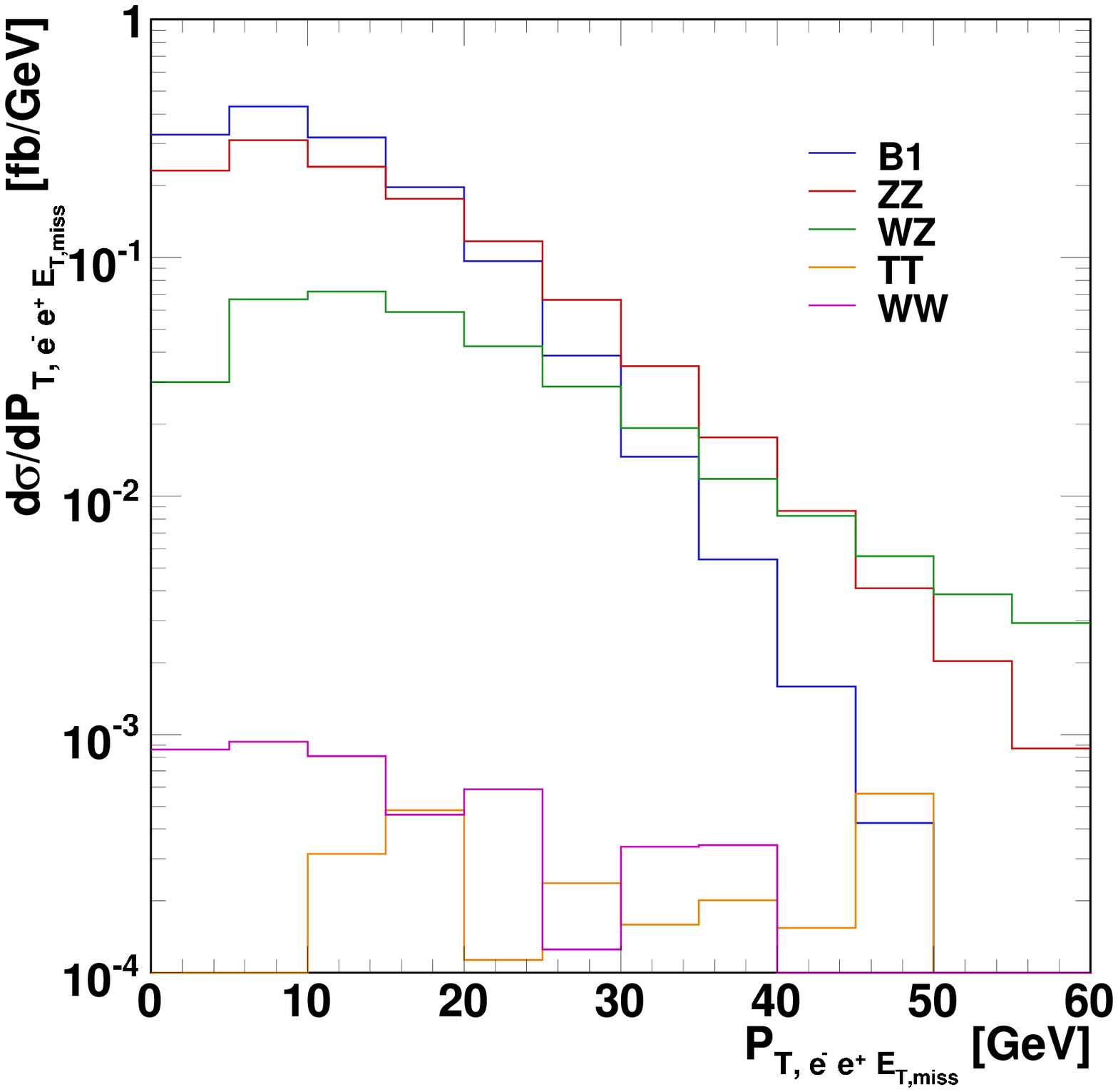}
\includegraphics[width=0.49\textwidth]{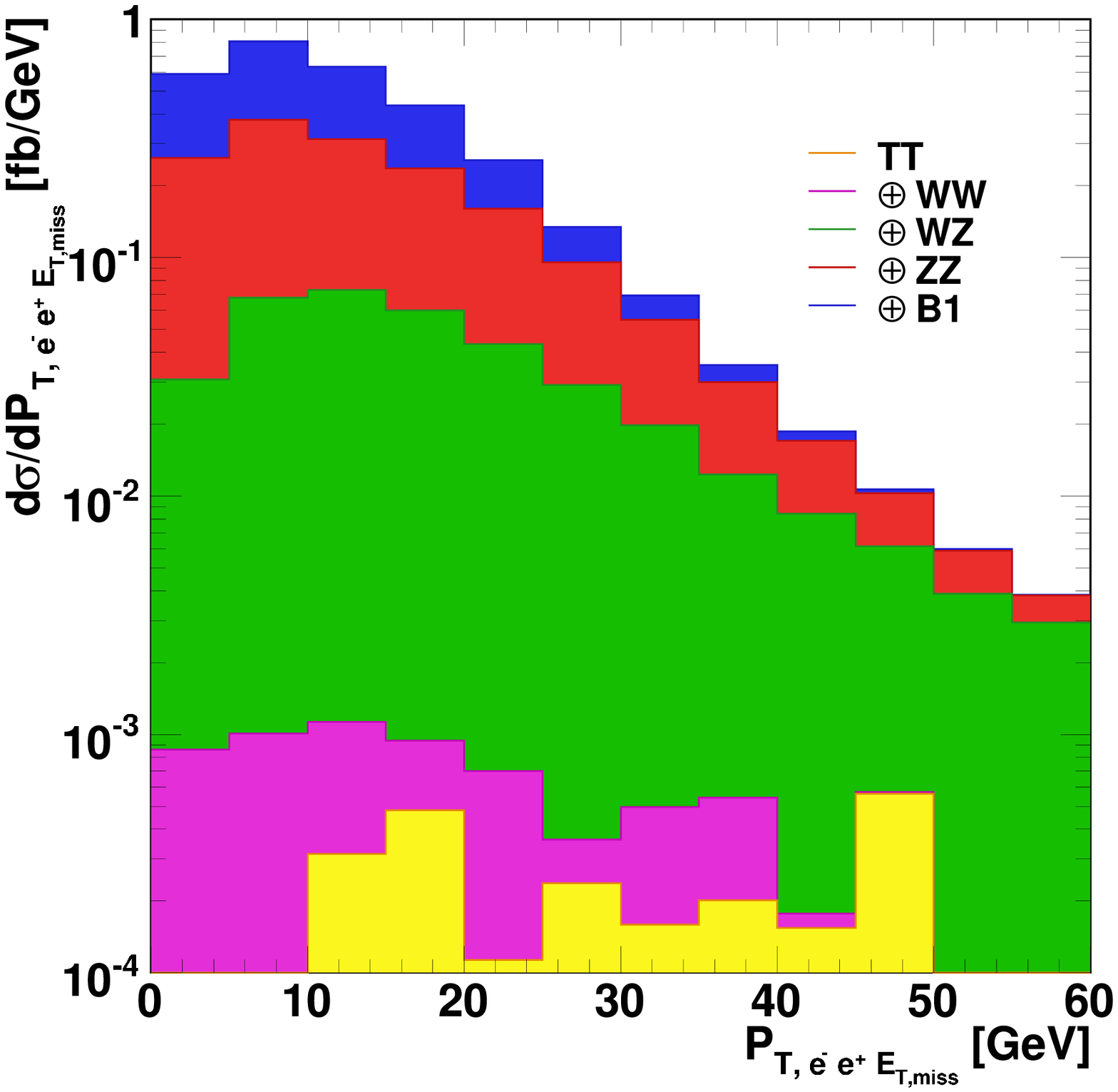}
\parbox{0.9\textwidth}{\caption{The $p_{T}(e^+e^-,\ETmiss)$ distribution for the signal 
	in benchmark scenario B1 and the individual backgrounds.  The left panel displays 
	individual distributions while the right panel show the sum of backgrounds and 
	signal, starting from the lowest significant background.} 
\label{fig:zh_plots1}}
\end{figure}

\begin{figure}[t]
\centering
\vspace{5mm}
\includegraphics[width=0.49\textwidth]{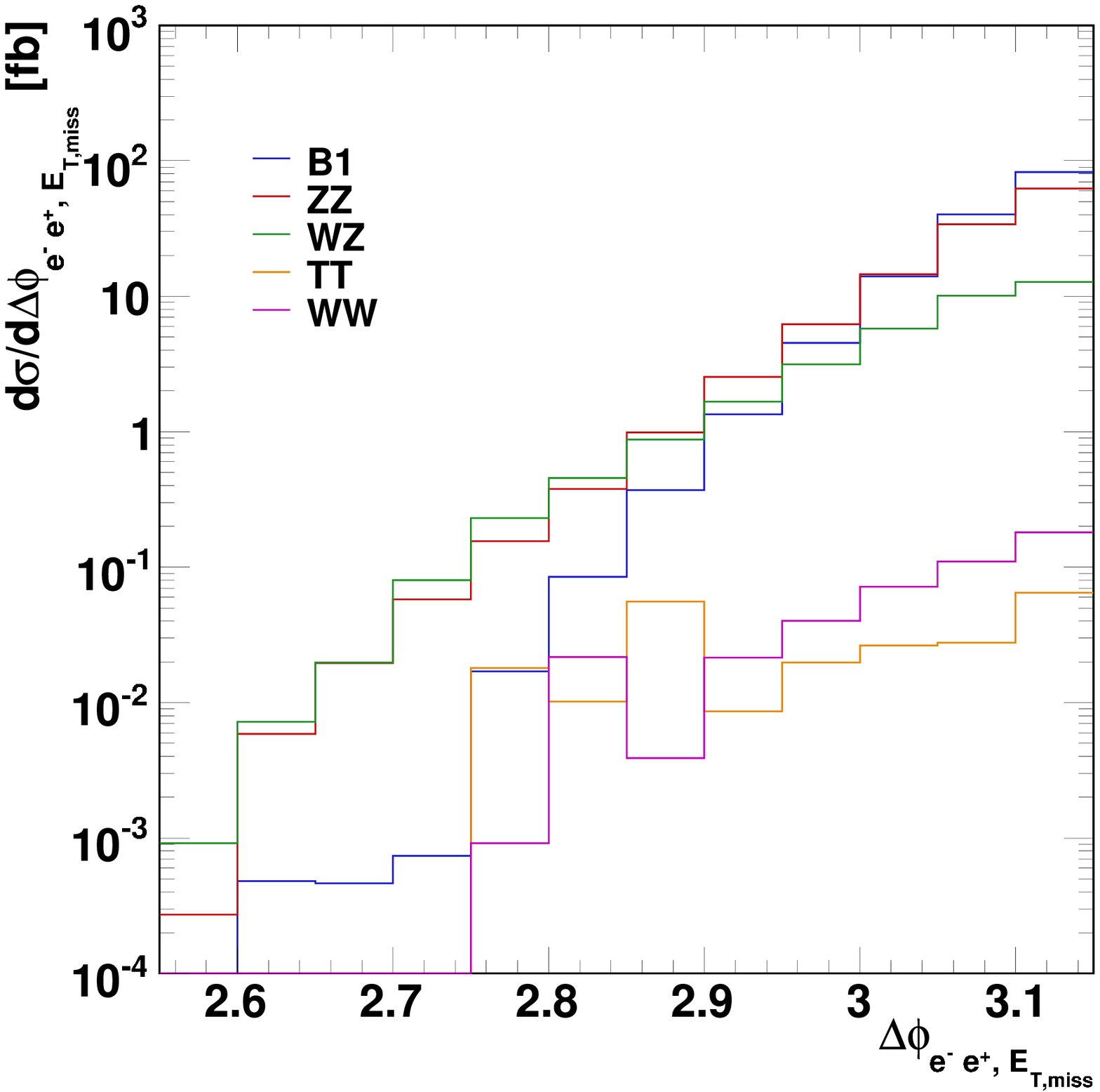}
\includegraphics[width=0.49\textwidth]{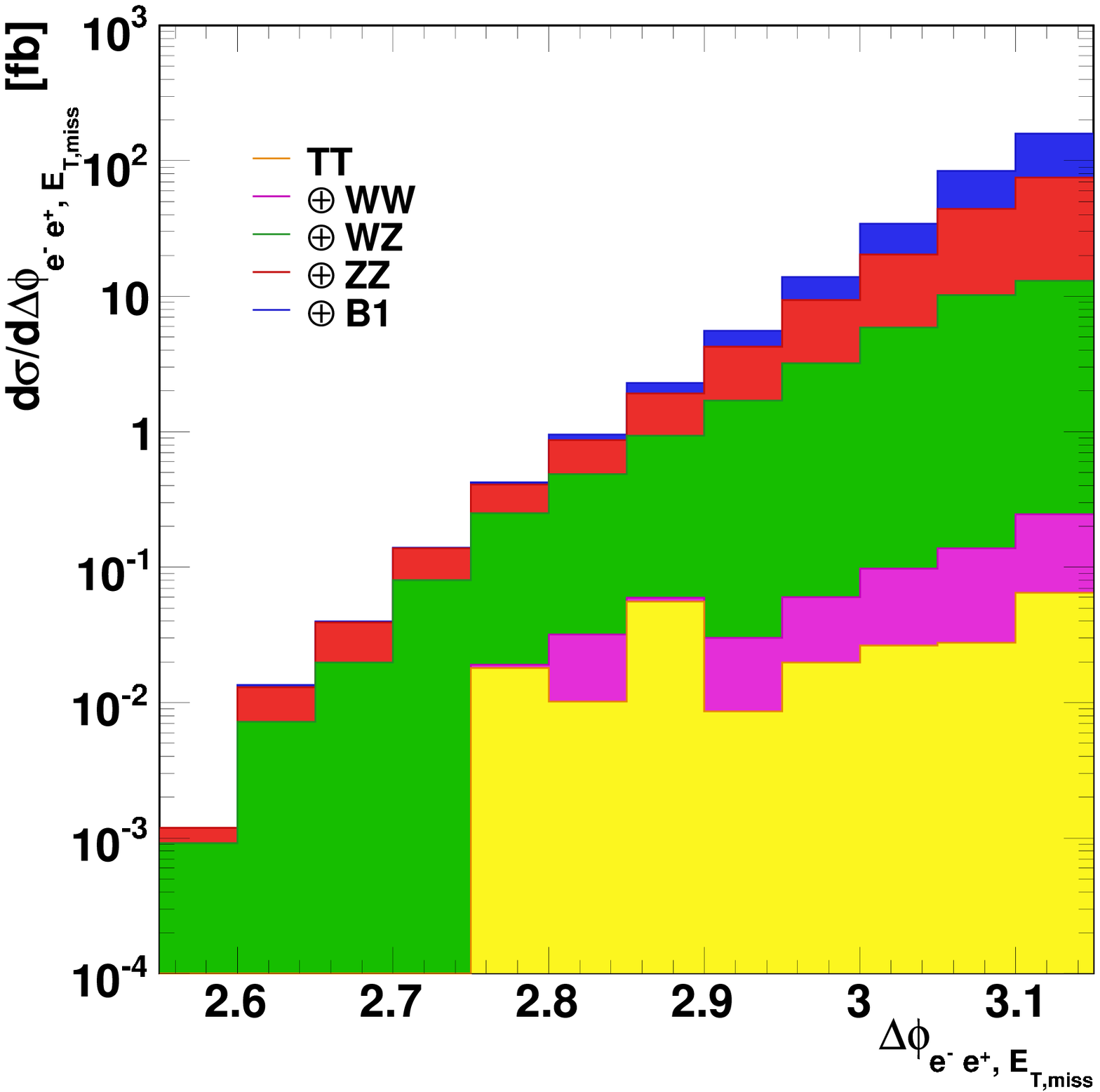}\\
\parbox{0.9\textwidth}{\caption{The $\Delta \phi(e^+e^-,\ETmiss)$ distribution for the 
	signal in benchmark scenario B1 and backgrounds.  The left panel displays individual 
	distributions while the right panel show the sum of backgrounds and signal, starting 
	from the lowest significant background.}\label{fig:zh_plots2}}
\end{figure}

In addition to the findings above, cf.\ Tables \ref{Tab:ZHbackground} and \ref{Tab:ZHsignal}
we have identified two further distributions that may be worthwhile to study in the $ZH$ 
channel:
\begin{itemize}
\item The total transverse momentum of the leptons and $\pTmiss$, i.e.\ the total transverse 
  momentum of the $H$ and $Z$ candidates (see Fig.\ \ref{fig:zh_plots1}).  This observable 
  shows a significantly different behaviour between the signals and the backgrounds, 
  where the signal tends to be more strongly peaked towards small values.
\item The azimuthal angle between $\pTmiss$ and	the momentum of the lepton pair
  (see Fig.\ \ref{fig:zh_plots2}).  Here the signal tends towards a more back-to-back 
  configuration of the $Z$ and $H$   candidate. Seemingly, there is a significantly higher 
  QCD activity in the backgrounds than in the signal, providing more jets for the 
  $ZH$-candidate pair to recoil against in the backgrounds.  
\end{itemize}
These findings may help to further improve the signal-to-background ratio.

\subsection{Vector Boson Fusion}
\label{VBFprod}

The other production channel we consider for invisibly decaying Higgs bosons at
the LHC is vector boson fusion (VBF).  As the name suggests, in this process the 
Higgs boson is produced through the fusion of two vector bosons emitted by quarks
inside the protons, which typically carry comparably large momentum fractions of
the protons.  Therefore, at leading order (tree-level) there is no colour exchange 
between the two protons, and it can be expected that the central rapidity region
remains to a large extent empty apart from the decay products of the produced
system.  The quarks on the other hand will be deflected, typically by transverse 
momenta of roughly half the mass of the produced system.  This gives rise to two 
hard jets, which, due to the invisible nature of the Higgs boson, are essentially 
the triggers in this analysis.  The main background processes to be taken into account 
are the production of $Z$ or $W$ bosons in association with two jets, which can
originate either from QCD or through electroweak interactions, thus mimicking the
topology of the VBF signal.  In addition, top-pair production with
subsequent semi-leptonic decays must be considered. Similar to the case of
$W$ production, the lepton is then lost.  Again, it is worth noting that it
should be possible to extract information concerning the total rates of these 
backgrounds, even after selection cuts, directly from data.  This is possible
either by reweighting leptonic $Z$ decays to those into neutrino pairs, or, with
a somewhat larger error, by extrapolating the modes where the individual lepton is
seen (in $W$+jets or semileptonic top-pairs) into those regions where the lepton
is lost.  This is in analogy to the case discussed above.
\begin{figure}[t]
\centering
\includegraphics[width=3.5in]{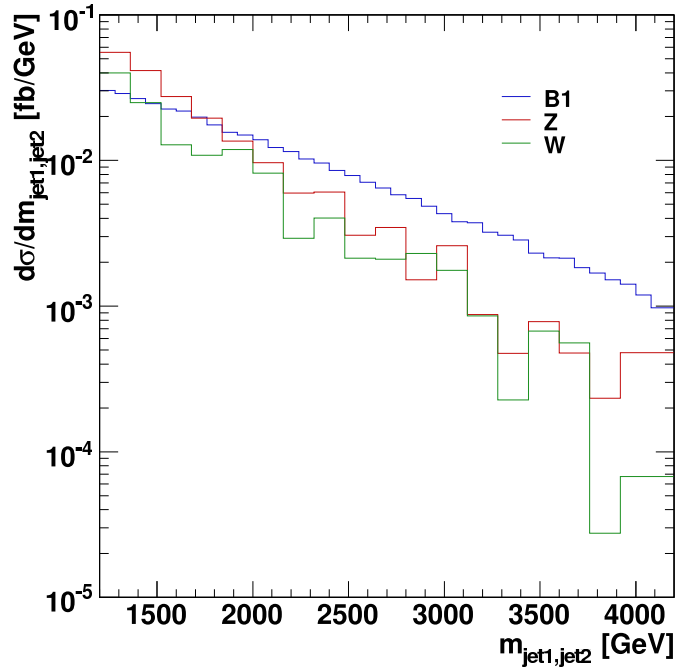}
\parbox{0.9\textwidth}{\caption{Tagjet invariant mass distribution
	for benchmark point B1 and the $Z$ and $W$ backgrounds.}
	\label{fig:mjj}}
\end{figure}
We employ the basic cuts listed in Ref.~\cite{VBFinv}, i.e.\ we require:
\begin{enumerate}
\item two tagging jets with 
	\begin{enumerate}
	\item $p_{T,j} > 40$ GeV, $|\eta_{j}|<5$,
	\item $|\eta_{j_1}-\eta_{j_2}|>5$, $\eta_{j_1} \cdot \eta_{j_2} < 0$,
	\item $m_{j_1 j_2} > 1700$ GeV ,
	\item $\Delta\phi_{j_1 j_2} = |\phi_{j_1} - \phi_{j_2}| < 1$,
  	\end{enumerate}
\item missing transverse momentum, $\pTmiss>100$ GeV;
\item no identified lepton, i.e.\ no lepton with\\
$p_{T}^{e,\mu} > 5\;, 6$ GeV in $|\eta_{l}| < 2.5$,
\item a central jet veto, i.e.\ no jets with\\ $p_T>20$ GeV,
	${\rm min}\{\eta_{j_1},\eta_{j_2}\}<\eta<{\rm max}\{\eta_{j_1},\eta_{j_2}\}$.
\end{enumerate}
Additionally we impose:
\begin{enumerate}
\item[5.] $|\eta_3^*| = 
	\left|\eta_{j_3} - \frac{1}{2}\left(\eta_{j_1}+\eta_{j_2}\right)\right| > 1.5$,
\item[6.] $\Delta\phi_{j_1,j_3},\;\Delta\phi_{j_2,j_3}\; < 1.25$.
\end{enumerate}
%
%
\begin{figure}[t]
\centering
\includegraphics[width=0.47\textwidth]{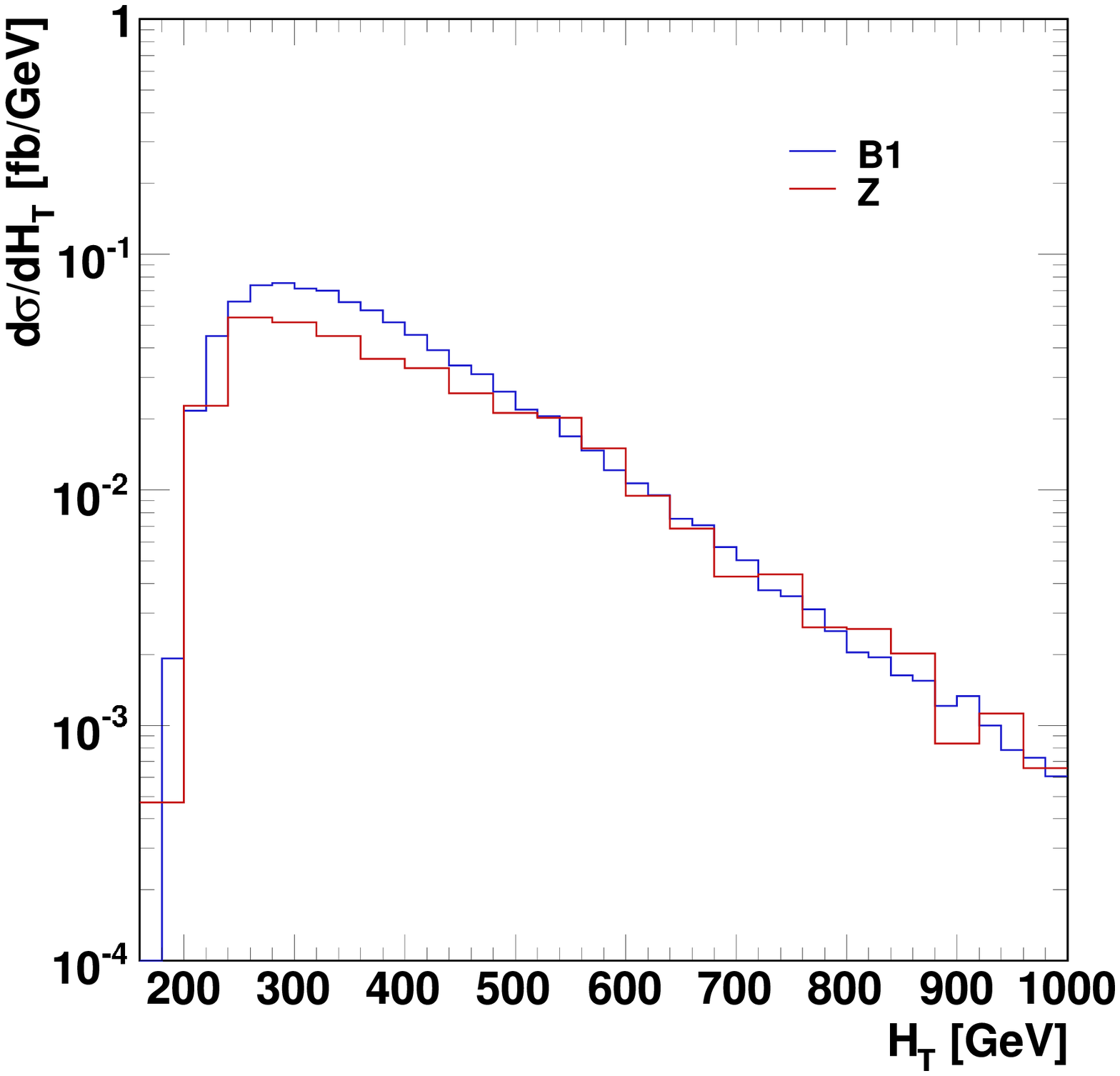} \hspace{3mm}
\includegraphics[width=0.46\textwidth]{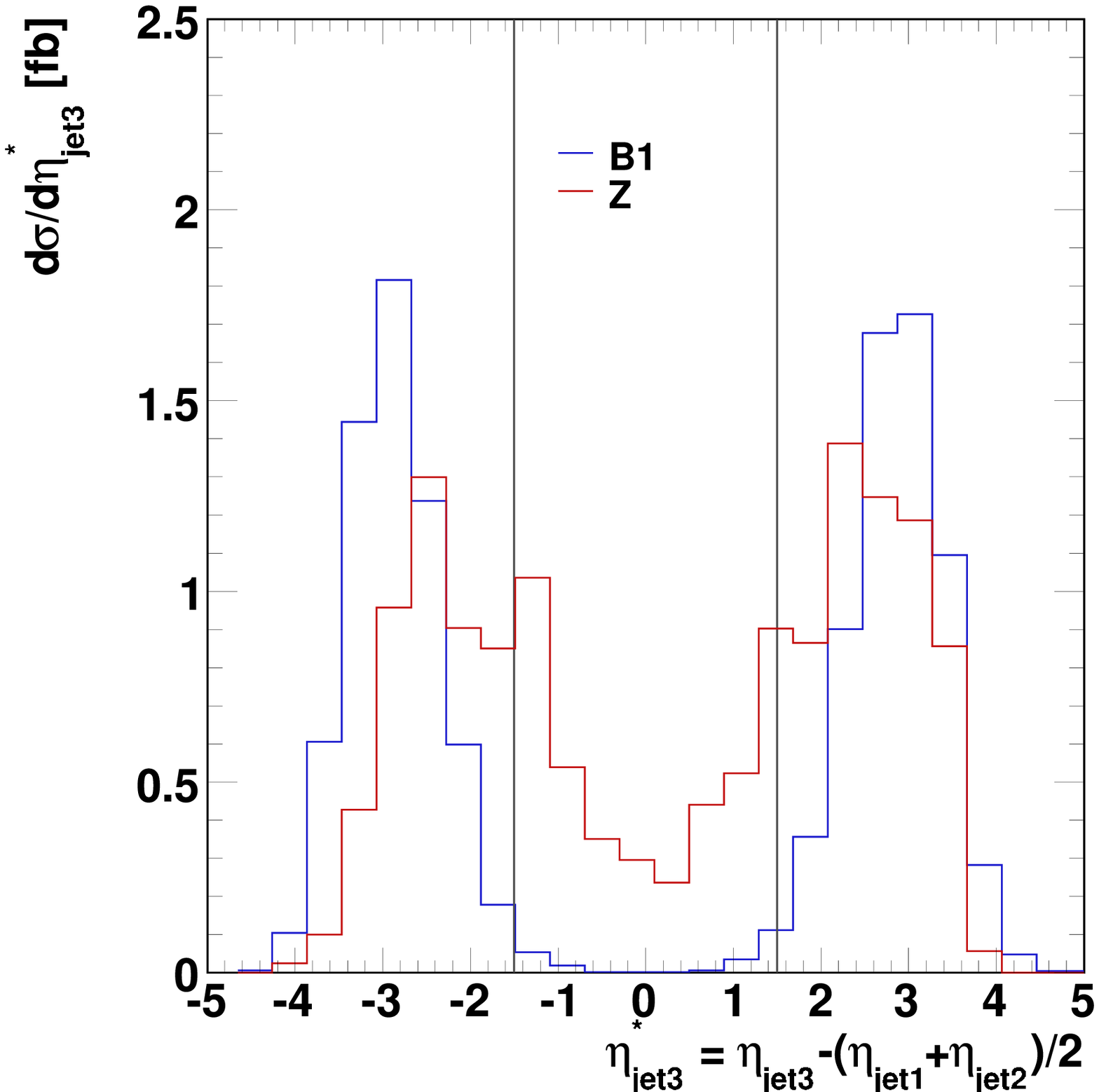}\\
\parbox{0.9\textwidth}{\caption{Left panel: $H_{T}$ distribution for benchmark point B1 
	and $Z$ background.  Right panel: $\eta^{*}_{3}$- distribution for benchmark 
	point B1 and $Z$ background.}
	\label{fig:eta3}}
\end{figure}

The choice of the cut on the invariant tagging jet mass of $m_{jj}> 1700$ GeV is 
motivated by the corresponding invariant mass spectrum shown in Fig.\ \ref{fig:mjj}.  
We observe that the signal distribution crosses the background at $m_{jj}\approx 1700$ GeV.  
Of course this statement sensitively depends on the model parameters chosen; however, the 
common feature of all scenarios is that the higher the invariant mass cut, the better the 
signal--\-to--\-bachground ratio.  This is due to the fact that in a large fraction of 
background events the two tagging jets originate from QCD or the decay of weak gauge 
bosons.

After the above cuts the possibilities to check for the signal topology are limited 
in the VBF channel.  Possible objects to be identified experimentally are the tagging 
jets, $\pTmiss$ and an eventually arising soft third jet. Therefore most observables 
show the same behaviour for signal and background, which is exemplified in the left 
panel of Fig.\ \ref{fig:eta3}, showing the $H_T$-distribution for the signal at 
benchmark point B1 and the $Z$ background.  In the right panel of Fig.\ \ref{fig:eta3} 
we show for the same scenario the $\eta^{*}_{3}$ distribution.   It is clearly 
seen that for the background the third jet tends to be more central between the 
tagging jets, while for the signal it is rather forward or backward.  This motivates 
the first of the additional cuts above.      

For the various backgrounds listed above, cross sections before and after additional 
selection cuts, and the number of generated events are listed in 
Table~\ref{Tab:VBFbackground}.  Signal cross sections before and after additional 
selection cuts are listed in Table~\ref{Tab:VBFsignal}.
\begin{table*}
\begin{center}
\begin{tabular}{|l||c|c|c|}
\hline
& \parbox{25mm}{\vspace{1mm}$Z$+jets (QCD+EW)\vspace{1mm}} 
& \parbox{25mm}{$W$+jets (QCD+EW)}
& \parbox{25mm}{$t\bar t$}\\\hline\hline
$\sigma_{\rm tot}^{\rm gen}$ [nb] & 9.41 & 51.8 & 0.145\\\hline\hline
tagging jets        & 1.80$\cdot 10^{-4}$ & 7.44$\cdot 10^{-5}$ & 1.62$\cdot 10^{-3}$ \\
$m_{jj}>$1700 GeV   & 3.49$\cdot 10^{-5}$ & 1.64$\cdot 10^{-5}$ & 4.44$\cdot 10^{-4}$ \\
$\pTmiss>100$ GeV   & 2.64$\cdot 10^{-5}$ & 9.73$\cdot 10^{-6}$ & 3.32$\cdot 10^{-4}$ \\ 
lepton veto         & 2.63$\cdot 10^{-5}$ & 2.84$\cdot 10^{-6}$ & 1.28$\cdot 10^{-5}$ \\ 
$\Delta\phi_{j_1j_2}<$1 & 4.03$\cdot 10^{-6}$ & 9.87$\cdot 10^{-7}$ & 2.79$\cdot 10^{-5}$ \\
central jet veto    & 1.54$\cdot 10^{-6}$ & 2.18$\cdot 10^{-7}$ & 1.44$\cdot 10^{-6}$ \\
$|\eta^*|>1.5$      & 1.37$\cdot 10^{-6}$ & 1.95$\cdot 10^{-7}$ & 6.70$\cdot 10^{-7}$ \\
$\Delta\phi_{j_1j_3}$, $\Delta\phi_{j_2j_3}$   
                    & 1.14$\cdot 10^{-6}$ & 1.44$\cdot 10^{-7}$ & 4.29$\cdot 10^{-7}$ \\
\hline\hline
$\sigma_{\rm eff}$ [fb] & 10.7 & 7.45 & 0.0621
\\\hline
\end{tabular}
\parbox{0.9\textwidth}{\caption{Generation characteristics for the background processes to 
	the VBF-channel.  Here, the $Z$ boson decays to neutrinos, whereas the $W$ boson 
	decays to any lepton--\-neutrino pair.  For the top--\-pairs, semileptonic decays 
	only have been considered.}
	\label{Tab:VBFbackground}}
\end{center}
\end{table*}
\begin{table*}
\begin{center}
\begin{tabular}{|l||c|c|c|c|c|}
\hline
                     & $B_1$   & $B_2$   & $B_3$   & $B_4$   & $B_5$   \\\hline\hline
$\sigma_{\rm tot}^{\rm gen}$ [pb] & 5.46 & 4.46 & 2.99 & 2.06 & 1.32 
\\\hline\hline
tagging jets        & 4.38$\cdot 10^{-2}$ & 5.59$\cdot 10^{-2}$ & 6.80$\cdot 10^{-1}$ & 7.54$\cdot 10^{-2}$ & 8.12$\cdot 10^{-1}$ \\
$m_{jj}>$1700 GeV   & 1.69$\cdot 10^{-2}$ & 2.20$\cdot 10^{-2}$ & 2.74$\cdot 10^{-2}$ & 3.07$\cdot 10^{-2}$ & 3.36$\cdot 10^{-2}$ \\
$\pTmiss>100$ GeV   & 1.46$\cdot 10^{-2}$ & 1.90$\cdot 10^{-2}$ & 2.37$\cdot 10^{-2}$ & 2.67$\cdot 10^{-2}$ & 2.92$\cdot 10^{-2}$ \\
lepton veto         & 1.46$\cdot 10^{-2}$ & 1.90$\cdot 10^{-2}$ & 2.37$\cdot 10^{-2}$ & 2.66$\cdot 10^{-2}$ & 2.92$\cdot 10^{-2}$ \\
$\Delta\phi_{j_1j_2}<$1 & 5.76$\cdot 10^{-3}$ & 7.65$\cdot 10^{-3}$ & 9.46$\cdot 10^{-3}$ & 1.06$\cdot 10^{-2}$ & 1.18$\cdot 10^{-2}$ \\
central jet veto    & 3.42$\cdot 10^{-3}$ & 4.33$\cdot 10^{-3}$ & 5.35$\cdot 10^{-3}$ & 6.06$\cdot 10^{-3}$ & 6.64$\cdot 10^{-3}$ \\
$|\eta^*|>1.5$      & 3.40$\cdot 10^{-3}$ & 4.31$\cdot 10^{-3}$ & 5.32$\cdot 10^{-3}$ & 6.03$\cdot 10^{-3}$ & 6.60$\cdot 10^{-3}$ \\
$\Delta\phi_{j_1j_3}$, $\Delta\phi_{j_2j_3}$   
                    & 3.11$\cdot 10^{-3}$ & 3.92$\cdot 10^{-3}$ & 4.81$\cdot 10^{-3}$ & 5.45$\cdot 10^{-3}$ & 5.97$\cdot 10^{-3}$ \\
\hline\hline
$\sigma_{\rm eff}$ [fb] & 17.0 & 17.5 & 14.4 & 11.2 & 7.9
\\\hline
\end{tabular}
\parbox{0.9\textwidth}{\caption{Generation characteristics for the signal processes in the 
	VBF-channel, for the different benchmark scenarios.}
	\label{Tab:VBFsignal}}
\end{center}
\end{table*}
Putting together numbers, we again find appreciable signal--\-to--\-background ratios between
more than 1/3 up to nearly 1 for all the benchmark points in the model.  However, this finding 
has to be taken with more than a pinch of salt: first of all, similar to the $ZH$ channel, we 
included all effects due to fragmentation, hadron decays, QED bremsstrahlung etc., and we 
typically added at least one further jet for a better modelling of additional hard QCD radiation.  
We did not, however, include the effects of the underlying event, which here could play a 
significant role in filling the rapidity gap between the two taging jets, and thus lead to a 
corresponding reduction in the effective cross section after selection cuts.  In addition we 
did not include diagrams where the Higgs boson is produced through an effective $ggH$ coupling, 
mediated by heavy quarks.  Although in principle the cross section for this mode is large, we 
note that previous work in the framework of the Standard Model suggests that the typical VBF 
cuts render this contribution insignificant 
\cite{DelDuca:2001fn,DelDuca:2001eu,Andersen:2007mp,Bredenstein:2008tm}.
Also, again, we did not simulate events at the detector-level which could further modify our 
findings.

However, again our results are in qualitative agreement with results of such a simulation at 
the detector level, which has been performed for the ATLAS experiment \cite{VBFinv}.  The 
results of this study were obtained using a fast detector simulation, and they are
quite encouraging, too.   Although in qualititative agreement, there are several 
differences: Again, the first one lies in the choice of the evenet generator.  ATLAS chose
\Pythia  to compute both signal and SM backgrounds at leading order, while we employed 
\Sherpa.  In the ATLAS simulation SM coupling strength for the $HVV$ couplings has
been assumed with a $100 \%$ branching fraction of the the Higgs decay to invisible, while in
our study the $HVV$ coupling is shielded through the mixing of the scalars, and the relevant
branching ratio ranges between 0.5 and 1.  While in ATLAS' \Pythia simulation the effect of
hard QCD radiation is typically accounted for by the parton shower, \Sherpa uses exact
matrix element, leading to a significantly increased jet activity.  Also, \Sherpa naturally
includes spin correlations, and VBF-like background topologies are also taken care off,
which have been missed in the ATLAS simulation.  These effects, together, would typically
reduce the signal--\-to--\-background ratio in our simulation with respect to the ATLAS 
study.  On the other hand, the effect of the underlying event as well as the fast detector
simulation, both included in the ATLAS analysis but ignored by us, may have the opposite
implications on the visibility of the signal.  Finally, it is worth stressing that we
have also chosen different optimization cuts, in particular cuts 5 and 6, to enhance the
signal over the background.  

Nevertheless, to summarize, we again find that the prospects of finding an invisibly decaying 
Higgs boson at the LHC are much better than naively anticipated, and the two channels 
considered here may very well play a significant role in the phenomenology of non-standard 
scalar sectors.

\section{Non-Abelian Phantom Sector}
\label{sec:nona}

So far only a $G_P = U(1)$ group theoretic phantom sector has been considered.  The
obvious question to be asked is how the Higgs boson observability will be affected 
in the case of non-Abelian extensions of the phantom sector (like $G_P = SU(N)$).  This will 
briefly be discussed in this section.  As an overall result, in general, such extensions 
typically result in further suppression of the Higgs boson visible event rates, 
$\mathcal{R}_i^2$.  Furthermore, in the case of more involved representations or multiple 
vector representations of $G_P$ the ``Higgs $\rightarrow$ invisible'' signal is decreased to 
a non-detectable rate.  Some examples supporting this result will be presented in the 
following.

Consider for instance a $G_P = SU(N)$ vector representation of scalar phantom
fields, $\vec{\Phi}$.  Then $SU(N)$ is spontaneously broken down to $SU(N-1)$
with $2N-1$ physical NGBs and one physical SM-singlet scalar field that
eventually mixes with the $SU(2)_L$ Higgs field. It is a textbook exercise to
prove that \eq{eq5} in such a framework becomes
\begin{eqnarray}
\mathcal{L}_{\rm int} \ = \ - \frac{m_{H_{i}}^{2}}{2 \,\sigma} \: O_{i2}\: H_{i}(x) \:
\mathcal{J}^a(x) \: \mathcal{J}^a(x) \qquad 
\mathrm{with} \quad a=1...(2N-1) \;.
\label{eq55}
\end{eqnarray}
This suggests that the Higgs boson decay width broadens compared to the $G_{P}=U(1)$
case.  The visible Higgs boson event rates (there are still two physical states) 
read
\begin{eqnarray}
{\cal R}^{2}_1 & \simeq & \Bigg[(1 + \tan^2 \theta)\Big(1 +
\frac{2N-1}{12}\,\frac{m_{H_{1}}^2}{m_b^2}\,\tan^2 \theta\,\tan^2 \beta\Big)\Bigg]^{-1}
 \;, \label{eqnonr1} \nonumber \\[3mm] 
{\cal R}^{2}_2 & \simeq & \Bigg[(1 + \cot^2 \theta)\Big(1 +
  \frac{2N-1}{12}\,\frac{m_{H_{2}}^2}{m_b^2}\,\cot^2 \theta\,\tan^2 \beta\Big)\Bigg]^{-1}
  \;. \nonumber  \label{eqrnon2}
\end{eqnarray}
Hence increasing the rank of the phantom gauge group results in a ($1/N$ for large $N$) 
decrease in visible Higgs boson rates.  Searching for ``Higgs $\rightarrow$ invisible'' is 
therefore vital.  Note also that increasing the rank of the phantom symmetry group does 
not necessarily imply different ``Higgs $\rightarrow$ invisible'' rates.  In fact, in the 
above example we still have two physical scalars in the spectrum for which the equation
$\mathcal{T}_1 + \mathcal{T}_2 \approx 1$ is valid, similarly to the $G_P = U(1)$ case.

It may also be the case that additional physical Higgs bosons fragment the
``Higgs $\rightarrow$ invisible'' rate into many small pieces such that any 
detection at the LHC seems completely impossible.  This case can be illustrated with 
the following example: consider $G_P= SU(3)$ broken by 2 sets of vector
representations down to the null group.  We start with 12 degrees of freedom,
out of which 8 become NGBs and the other 4 become massive scalar fields.  These 4
fields will mix with the one $SU(2)_L$ Higgs field through the ($5\times 5$)
matrix $O$ forming 5 physical Higgs-boson eigenstates. In this case, due to
the unitarity of the matrix $O$ we have $\sum_{i=1}^5 \mathcal{T}_i^2 \approx
1$, which allows for $\mathcal{T}_i^2 \lesssim 0.25$.  Such a ``Higgs
$\rightarrow$ invisible'' rate is most probably beyond reach of discovery (or exclusion) 
at the LHC~\cite{VBFinv,HZinv} - a truly nightmarish scenario!

\section{Additional Remarks}
\label{sec-remarks}

It should be emphasized that in the scenario considered in this article, invisible 
Higgs boson phenomenology, small neutrino masses and the correct baryon asymmetry (see 
also \Ref{CDU}) are all obtained without fine-tuning coupling constants.  All 
scalars have masses at the EW scale ($\tan\beta \approx 1$) and so there are no 
ultra-heavy scalars to destabilize this hierarchy.  However, the model does not 
include gravity nor does it contain a mechanism or theoretical explanation as to 
why $\sigma \ll M_{\rm Planck}$.  Although the SM hierarchy problem is not solved 
in this model the question here is somewhat different:
{\em Why is the phantom sector symmetry broken at the EW scale ?}
We cannot provide a non-common ({\it i.e.}, non-supersymmetric) answer to this question,
and refer to \cite{WilczekSUSY08,JungKo}. 

Instead of a theory with one global symmetry, one could imagine a theory where
several symmetries were gauged (or left un-gauged), absorbing the NGBs into
massive gauge bosons through the Higgs mechanism.  This is an absolutely viable
option, although the requirement of anomaly cancellation would result in model
dependencies.  Such models have been proposed before and studied in some detail
in the recent literature \cite{gaugemodels}.  Generally speaking, these models
lead to phenomenology that includes the (observable) decays of the extra gauge
bosons, with all constraints on their masses etc..
%

Recently there has been renewed interest in the possibilities offered by
extending the Standard Model with a real scalar singlet \cite{langackeretal}.
Depending on the symmetries of the model it is possible to provide a candidate
for the cold dark matter in the universe (extra discrete symmetries needed) 
\cite{mcdonald} , and it is possible to provide a strong first-order electroweak 
phase transition suitable for electroweak baryogenesis \cite{quirosespinosa}.  
It should be noted in the latter case that an additional source of CP-violation 
would be necessary to provide a complete mechanism for baryogenesis.

Models with broken discrete symmetries provide another possible way of
avoiding invisible decays of Higgs boson(s). Clearly, spontaneous breaking of
such symmetries does not lead to NGBs, making the Higgs boson signatures more
visible.  There are, however, so many possibilities of such groups that a
particular choice renders this idea less appealing and convincing.
Spontaneously broken discrete symmetries may also, in some cases, produce
unwanted cosmological relics such as domain-walls, potentially placing severe
constraints on this class of model.

\section{Conclusions}
\label{sec:conclusions}

Physical NGBs arise when continuous global symmetries are spontaneously
broken.  Such broken symmetries may be related to the smallness of neutrino
masses or the patterns of mixing angles (in the case of familons).  In this
article we show that the role of NGBs in Higgs boson phenomenology is very
important; they lead to the dilution and potential invisibility of the expected SM
signal.  Working with approximate analytic formulae we first identified
regions of parameter space [\eq{eq6}] where Higgs boson phenomenology is
challenging both for past LEP data and for the future LHC experiments, and
secondly implemented the model in \Sherpa, ready for further analysis by
experimenters when real LHC data arrive.

Our study shows that LEP excludes the minimal phantom sector case where both
Higgs bosons have masses $m_H \stackrel{<}{{}_\sim} 85$~GeV irrespective of
their decay modes. However, experimentally allowed scenarios exist where one
Higgs boson mass is much lower than the SM Higgs boson exclusion limit,
$m_{H_1} = 68$~GeV, and the other is just at this limit, $m_{H_2} = 114$~GeV.

In light of the nightmarish potential of this scenario, Monte-Carlo simulation
studies of invisible Higgs boson searches at the LHC are performed. Two search
channels are looked at in detail; the associated production of a $Z$ and a Higgs 
boson, and the production of a Higgs boson in weak vector boson fusion.
For $ZH$ associated production, it is found that in each of 5 benchmark scenarios, 
the invisible Higgs boson should be found at the LHC, with signal--\-to--\-background ratios 
of order $S/B \simeq 1/8$ to 1.  Scope for improving this ratio is also found by looking 
either at the distribution of the total transverse momentum of the leptons and the 
\pTmiss, or at the distribution of the azimuthal angle between the \pTmiss~and the
momentum of the lepton pair.
Fairly good signal--\-to--\-background ratios are also found in the vector boson fusion 
search channel.  However in this case the effects of the underlying
event, which was not included in simulations, may reduce the amount of signal
passing the selection cuts.

Although our MC analysis focuses on the case with an Abelian phantom sector
symmetry, we also examined cases with non-Abelian symmetries in the phantom
sector using the analytic formulae provided in section~\ref{Q1LEP}.  For the
case $G_P = SU(N)$ we found that the visibility of the Higgs bosons is reduced
when we increase the rank of the $SU(N)$ group making the LHC searches to
invisible a necessity.  In addition, by choosing appropriate representations of
the group for breaking the symmetry we may further dilute the Higgs boson to
invisible signature, leading to a very difficult scenario indeed for the LHC.

Regarding the hierarchy problem, the model at hand is not better or worse than
the Standard Model.  Any difference could be interpreted as shifting the
problem to the phantom sector which sets the scale of the symmetry breaking.

\vspace*{0.5cm}
\noindent {\large \bf Acknowledgements}

We would like to thank Karl Jacobs for letting us know about progress on ATLAS
studies of Higgs to invisible. A.D.\ would like to thank the European Artemis network 
and especially  Nikos Konstantinidis for discussions on ``Higgs to invisible'' at LEP.  
We are all fully (S.H.) or partially supported by the RTN European Programme, 
MRTN-CT-2006-035505 (HEPTOOLS, Tools and Precision Calculations for Physics Discoveries 
at Colliders) or by the RTN European Programme MRTN-CT-2006-035606 (MCnet).

 \renewcommand{\thesection}{Appendix~\Alph{section}}
 \renewcommand{\theequation}{\Alph{section}.\arabic{equation}}
 \setcounter{equation}{0}  
 \setcounter{section}{0}
 \bigskip

 
\section{$U(1)$ Phantom Model Feynman Rules}\label{app:A}

\begin{figure*}[t]
\begin{tabular}{cc}
\parbox{2cm}{\includegraphics[scale=0.3]{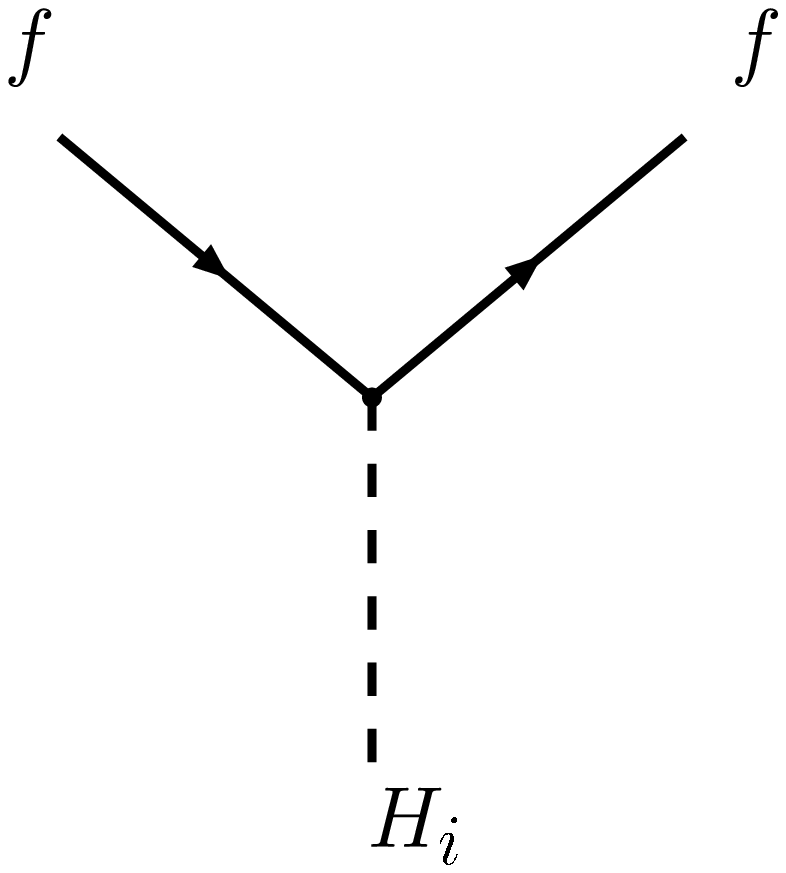}} &
\parbox{7cm}{: ~$-i \frac{m_{f}}{v} ~O_{i1}$} \\ \\
\parbox{2cm}{\includegraphics[scale=0.3]{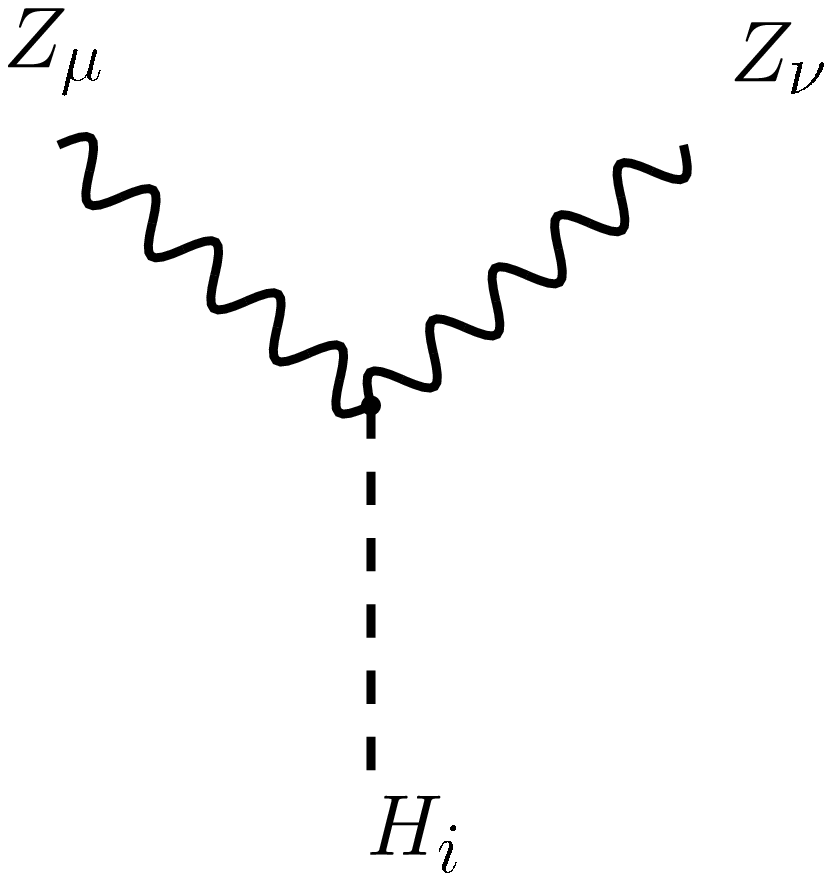}} &
\parbox{7cm}{: ~$i \frac{g_{2} M_{Z}}{\cos \theta_{w}} ~O_{i1} ~g_{\mu \nu}$}  \\ \\
\parbox{2cm}{\includegraphics[scale=0.3]{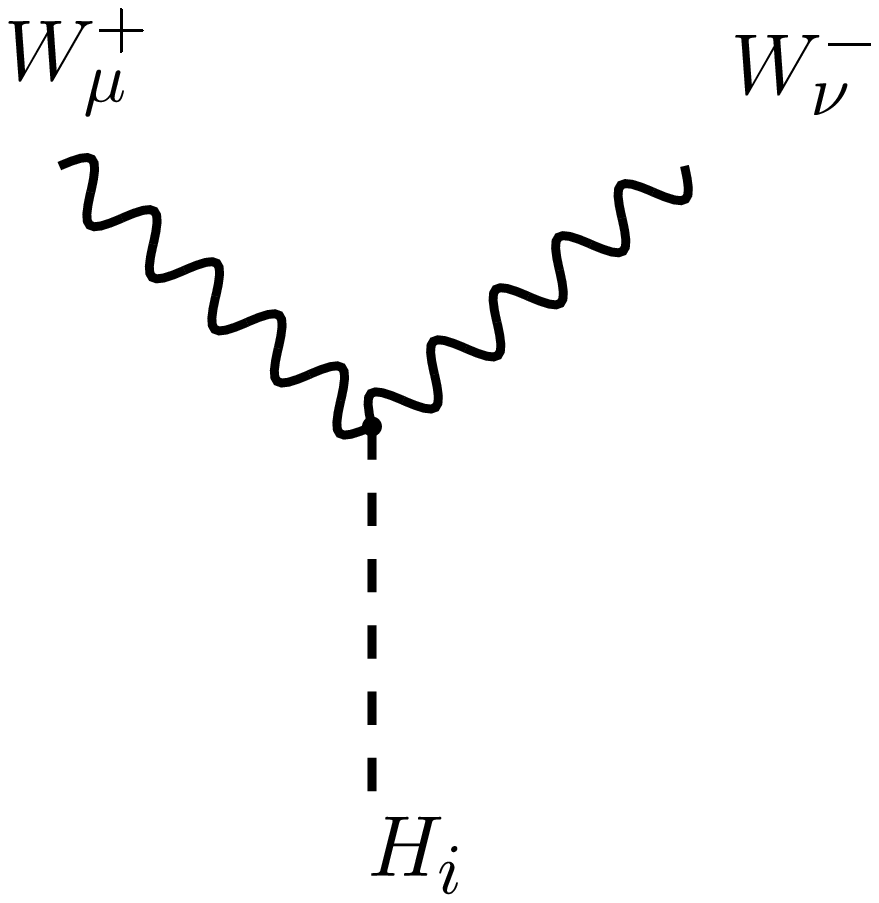}} &
\parbox{7cm}{: ~$i g_{2} M_{W} ~O_{i1} ~g_{\mu \nu}$} \\ \\
\parbox{2cm}{\includegraphics[scale=0.3]{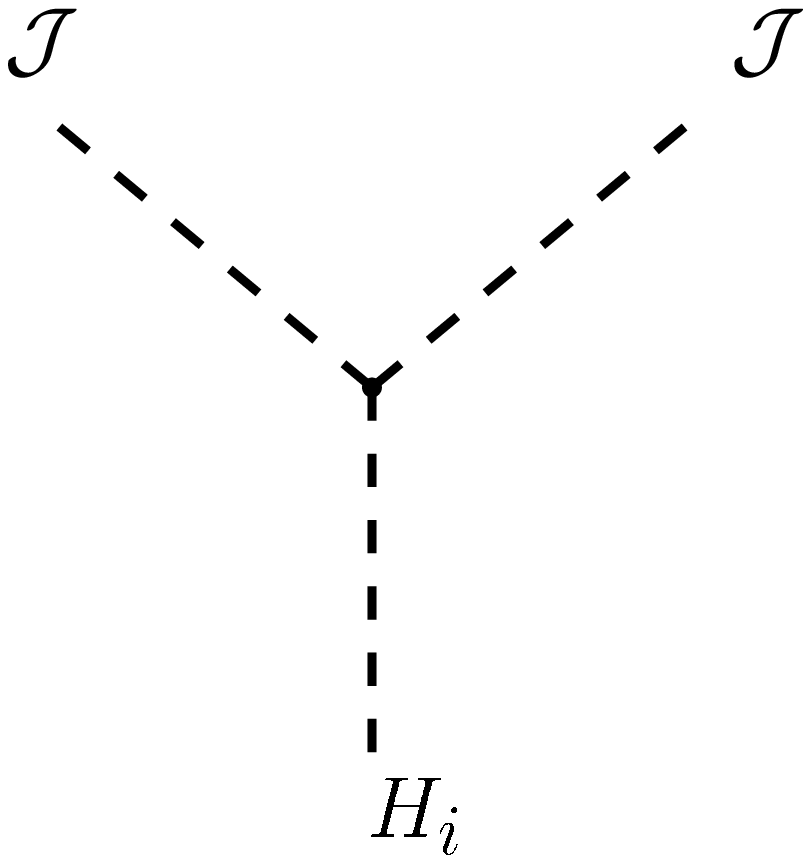}} &
\parbox{7cm}{: ~$-i \frac{m_{H_{i}}^{2}}{v} ~\tan \beta ~O_{i2}$} \\ \\
\parbox{2cm}{\includegraphics[scale=0.3]{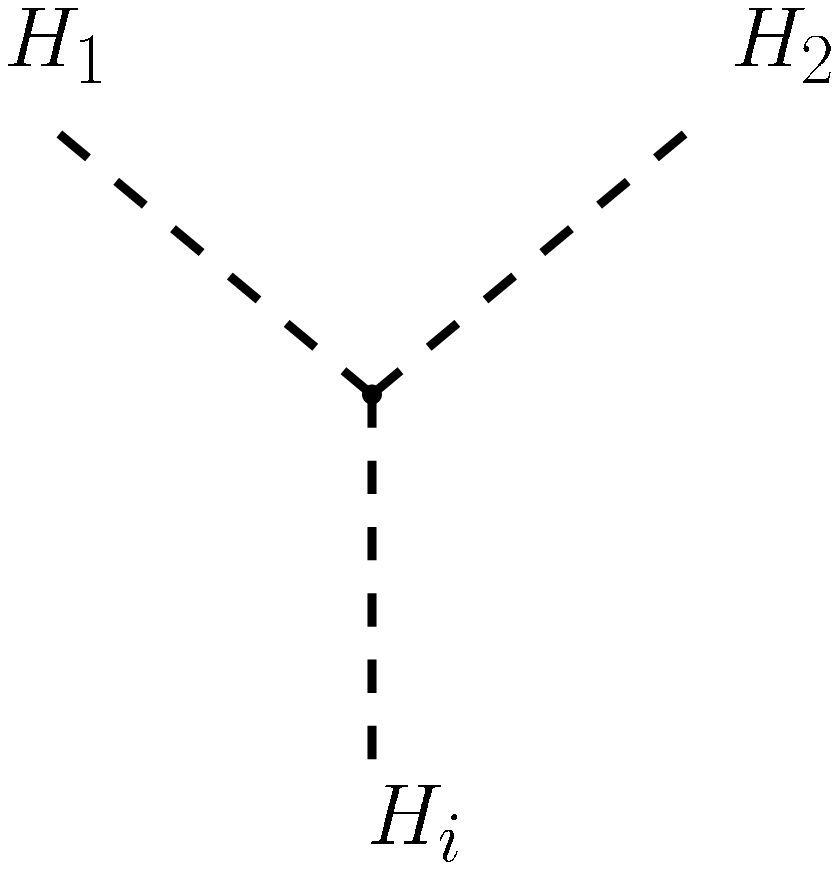}} &
\parbox{7cm}{\begin{eqnarray*}
: ~& & \frac{i}{v} \left( m_{H_{1}}^{2} + m_{H_{2}}^{2}+ m_{H_{i}}^{2} \right) \\
&\cdot & O_{i1} O_{i2} \left( O_{1i}+ O_{2i} ~\tan \beta \right)
\end{eqnarray*}}  \\ \\
\parbox{2cm}{\includegraphics[scale=0.3]{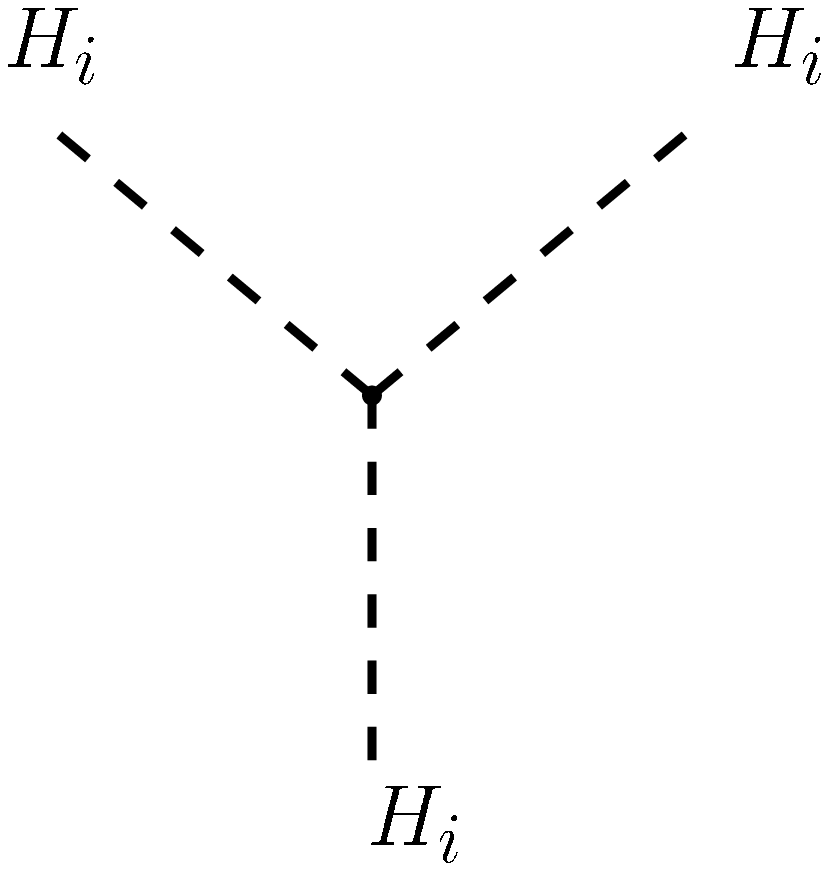}} &
\parbox{5cm}{: ~$-3i \frac{m_{H_{i}}^{2}}{v} ~\left(O_{i1}^{3}+O_{i2}^3 ~\tan \beta \right) $} \\ \\
\end{tabular}
\caption{\em Trilinear couplings.}
\label{triverts}
\end{figure*}

\begin{figure*}[t]
\begin{tabular}{cc}
\parbox{2cm}{\includegraphics[scale=0.3]{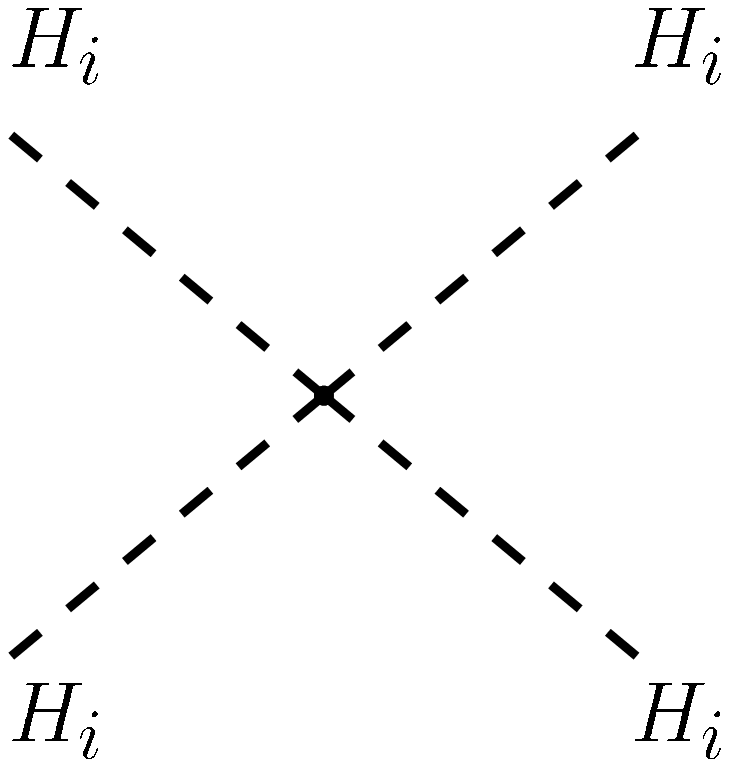}} &
\parbox{7cm}{\begin{eqnarray*}
:~&-&\frac{3i}{v^2} \left[ O_{1i}^{4} \left( m_{H_{1}}^{2} O_{11}^{2} + m_{H_{2}}^{2} O_{12}^{2} \right) \right. \\
&+& O_{i2}^{4} \left( m_{H_{2}}^{2} O_{11}^{2} + m_{H_{2}}^{2} O_{12}^{2} \right) \tan^{2} \beta \\
&-& \left. 2 O_{12}^{3} O_{11}^{3} \tan \beta \left( m_{H_{2}}^{2} - m_{H_{1}}^{2} \right) \right]
\end{eqnarray*} } \\ \\
\parbox{2cm}{\includegraphics[scale=0.3]{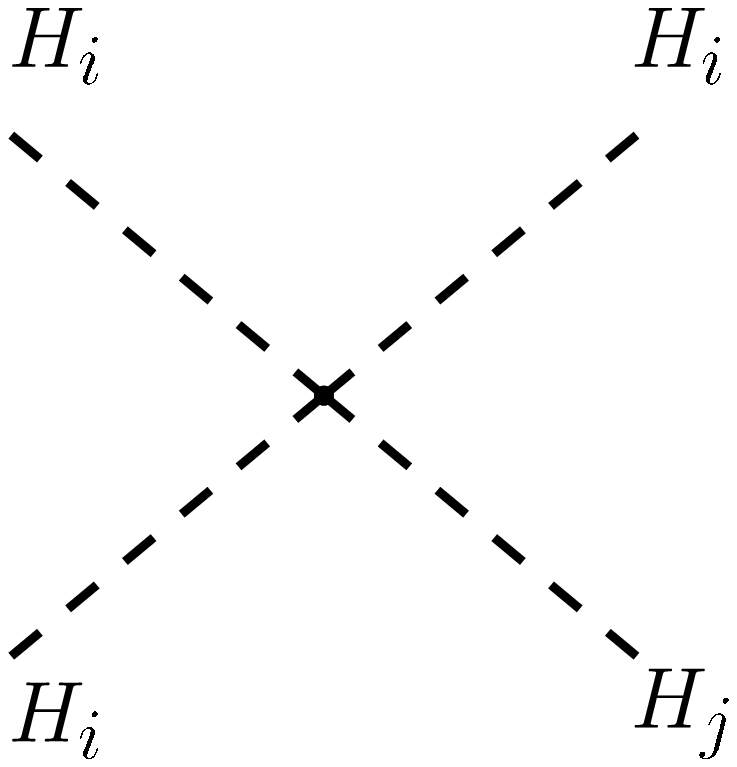} $(i \neq j) $} &
\parbox{7cm}{\begin{eqnarray*}
: ~&+&\frac{3i}{v^2} O_{11} O_{ij} \left( O_{j2}+O_{j1} \tan \beta \right) \\
&\cdot & \left[ m_{H_{i}}^{2} \left( O_{i2}^{3} \tan \beta + O_{i1}^{3} \right) \right. \\
&+& \left. m_{H_{j}}^{2} \left(O_{i1} O_{j1}^{2} + O_{i2} O_{j2}^{2} \tan \beta \right) \right]
\end{eqnarray*}
}  \\ \\
\parbox{2cm}{\includegraphics[scale=0.3]{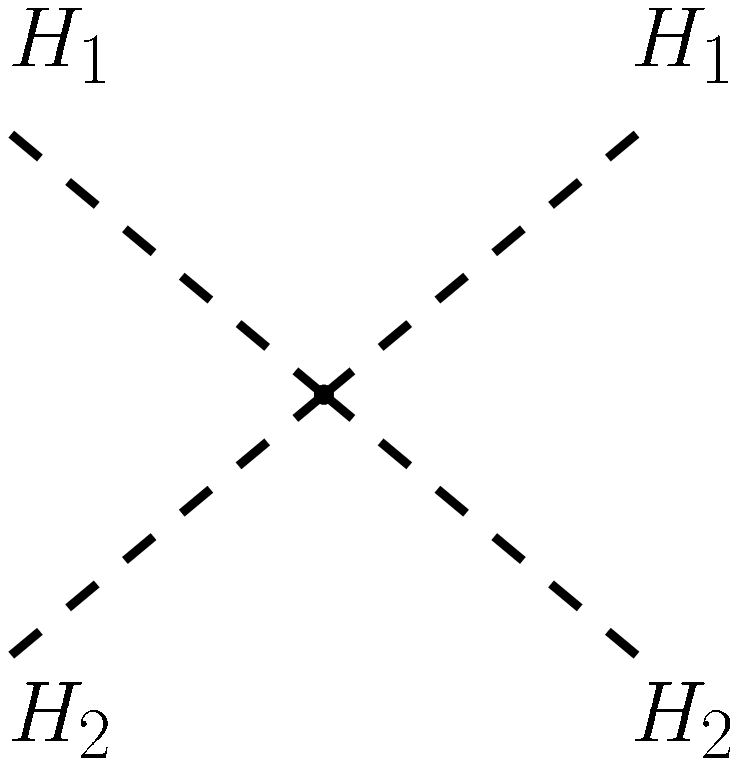}} &
\parbox{7cm}{\begin{eqnarray*}
:~&+&\frac{i}{v^2} O_{11} O_{12} \left[ \left( m_{H_{2}}^{2} - m_{H_{1}}^{2} \right) \tan \beta \right. \\
&\cdot& \left( O_{11}^{4} - 4 O_{12}^{2} O_{11}^{2}+O_{12}^{4} \right)  \\
&+& 3 O_{11} O_{12} \left( m_{H_{1}}^{2} O_{11}^{2} + m_{H_{2}}^{2} O_{12}^{2} \right.\\ 
&+&\left. \left. \left( m_{H_{2}}^{2} O_{11}^{2} + m_{H_{1}}^{2} O_{12}^{2} \right) \tan^{2} \beta \right) \right] 
\end{eqnarray*}} \\ \\
\parbox{2cm}{\includegraphics[scale=0.3]{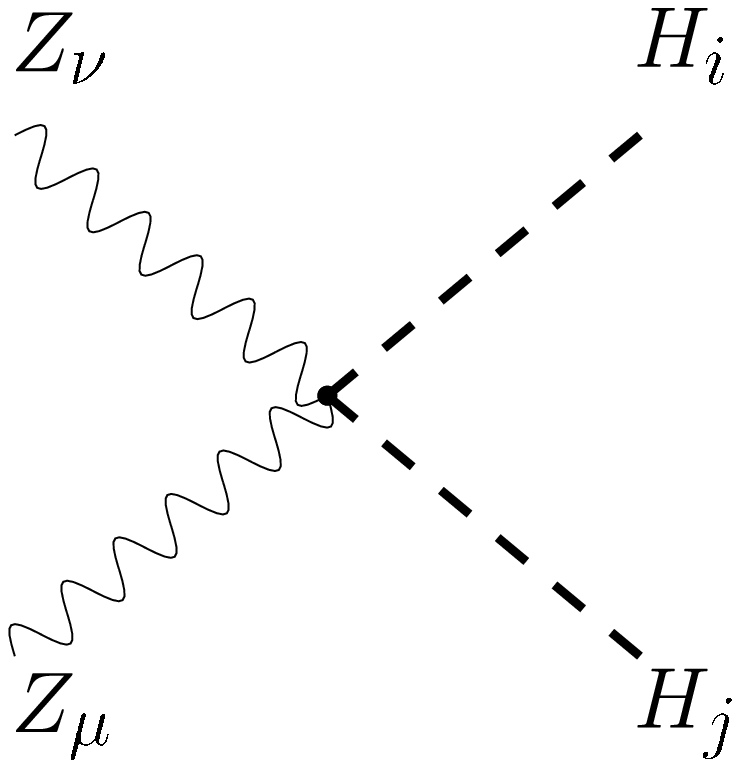}} &
\parbox{7cm}{: ~$\frac{ig_{2}^{2}}{2} g_{\mu \nu} O_{i1} O_{j1}$} \\ \\
\parbox{2cm}{\includegraphics[scale=0.3]{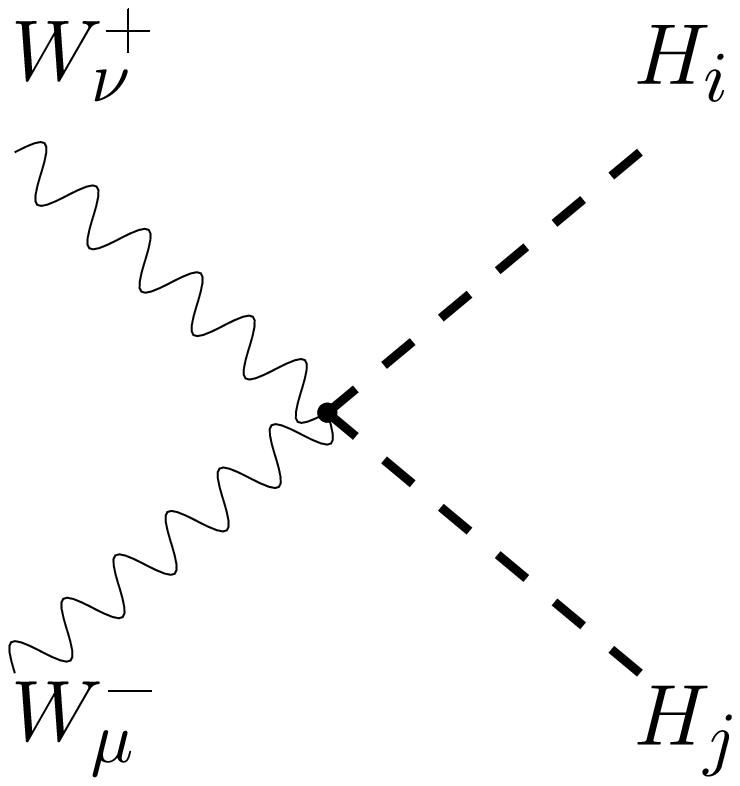}} &
\parbox{7cm}{: ~$\frac{ig_{2}^{2}}{2 \cos^{2} \theta_{w}} g_{\mu \nu} O_{i1} O_{j1} $}  \\ \\
\parbox{2cm}{\includegraphics[scale=0.3]{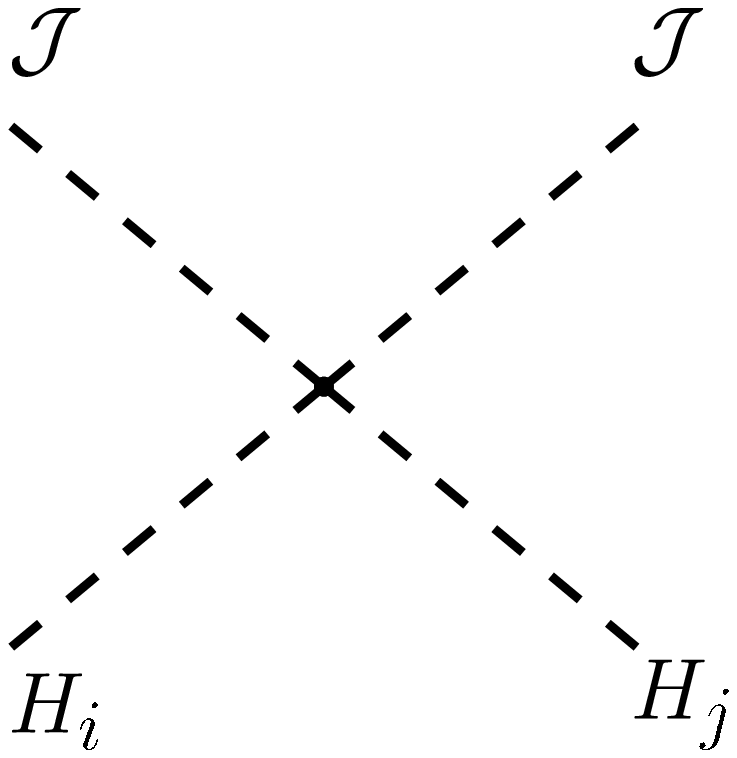}} &
\parbox{7cm}{\begin{eqnarray} \nonumber 
 :~ &+&\frac{i}{v^2}    \left[- O_{i2} O_{j2} \left(m_{H_{1}}^{2} O_{12}^{2} +
m_{H_{2}}^{2} O^{2}_{11} \right) ~\tan^2 \beta \right. \\ \nonumber 
&+&  \left. \left(m_{H_{2}}^{2} - m_{H_{1}}^{2} \right) O_{12} 
O_{11} O_{i1} O_{j1} ~\tan \beta  \right]
\end{eqnarray} } \\ \\

\end{tabular}
\caption{\em Quadrilinear couplings.}
\label{quadverts}
\end{figure*}

In this appendix we present Feynman rules for the Higgs sector of 
$G_{SM} \times \{G_{P}=U(1)_{P}\}$ that are relevant
 for Higgs phenomenology at LEP and the LHC.
Feynman rules for the trilinear couplings 
$H_{i} \mathcal{J} \mathcal{J}$, $H_{1}H_{2}H_{i}$, 
$H_{i}H_{i}H_{i}$, $W^{+}W^{-}H_{i}$, $ZZH_{i}$, 
and $f\bar{f}H_{i}$ for $i=1,2$ are shown in Fig.\ \ref{triverts}. 
For completeness in Fig.\ \ref{quadverts}, also Feynman rules for the quadrilinear 
couplings $H_{i}H_{j} \mathcal{J} \mathcal{J}$, $H_{i}H_{j}H_{k}H_{l}$, 
$H_{i}H_{j}ZZ$, and $H_{i}H_{j}W^{+}W^{-}$ are listed.

$M_{W}$  and $M_{Z}$ are the masses of the $W$ boson and $Z$ boson, respectively and
$m_{f}$ is the fermion mass which can be either a quark or a lepton.  The $SU(2)_{L}$ 
coupling constant is $g_{2}$ and $\theta_{w}$ is the Weinberg mixing angle.  $v$ is 
the vacuum expectation value for the standard model $SU(2)_{L}$ Higgs doublet $H$.  
$g_{\mu \nu}$ is the Minkowski spacetime metric $(1,-1,-1,-1)$.

The orthogonal mixing matrix, $O$, is 
\begin{eqnarray}
O=\left( \begin{array}{cc}
O_{11} & O_{12}  \\ 
O_{21} & O_{22} 
\end{array} \right) =
\left( \begin{array}{cc}
 \cos \theta & \sin \theta  \\ 
-\sin \theta & \cos \theta 
\end{array} \right) .
\end{eqnarray}
Here $ \tan \beta=v/\sigma$ with $\sigma  \equiv \langle \Phi \rangle $.
$m_{H_{1}}$ and $m_{H_{2}}$ denote the masses of the two Higgs bosons, $H_{1}$ and $H_{2}$, respectively.

\clearpage


\end{document}